\documentclass[review]{elsarticle}
\usepackage{hyperref}
\usepackage{float}
\usepackage{verbatim} 
\restylefloat{figure}
\restylefloat{table}
\usepackage[ruled,vlined]{algorithm2e}
\usepackage{algorithmic}
\usepackage{graphicx}
\usepackage{hyperref}
\usepackage{rotating}
\usepackage{amsmath}
\usepackage{xcolor}
\usepackage{amssymb}
\usepackage{amsfonts}
\usepackage{adjustbox}
\usepackage{tabularx}
\usepackage{url}
\newtheorem{Algoritmo}{\em Algorithm}
\usepackage{xurl}
\journal{Elsevier}
\journal{XXX}
\usepackage{hyperref}

\biboptions{authoryear}
\usepackage{color}
\usepackage{amsfonts,amssymb,amsmath}
\usepackage{hyperref}
\usepackage{url}
\usepackage{amsfonts}
\usepackage{graphicx}
\usepackage{amsmath}
\usepackage{graphicx}
\usepackage{array}
\usepackage{todonotes}
\usepackage{tabularx}
\usepackage{rotating}
\usepackage{amsmath}
\usepackage{xcolor}
\usepackage{amssymb}
\usepackage{amsfonts}
\usepackage{url}
\usepackage{appendix}
\usepackage{color}
\usepackage[utf8]{inputenc}
\usepackage[T1]{fontenc}
\usepackage{mathtools}
\usepackage{enumitem}
\usepackage{longtable}
\usepackage{multirow}
\def\A{{\mathbf A}}
\def\B{{\mathbf B}}
\def\C{{\mathbf C}}
\def\D{{\mathbf D}}

\def\R{{\mathbf R}}
\def\P{{\mathbf P}}
\def\K{{\mathbf K}}

\def\I{{\mathbf I}}

\def\Q{{\mathbf Q}}
\def\F{{\mathbf F}}

\def\G{{\mathbf G}}

\def\S{{\mathbf S}}
\def\T{{\mathbf T}}

\def\z{{\mathbf z}}
\def\v{{\mathbf v}}
\def\x{{\mathbf x}}

\def\y{{\mathbf y}}
\def\z{{\mathbf z}}

\def\u{{\mathbf u}}

\newcommand{\Sigmab}{\boldsymbol{\Sigma}}
\newcommand{\Phib}{\boldsymbol{\Phi}}
\newcommand{\Deltab}{\boldsymbol{\Delta}}
\newcommand{\cred}{\textcolor{black}}
\newcommand{\cblue}{\textcolor{black}}
\newcommand{\cmazenta}{\textcolor{black}}
\begin{document}
\begin{frontmatter}

\begin{titlepage}
\begin{center}
\vspace*{1cm}

\textbf{\cmazenta{Deep State-Space Model for Predicting Cryptocurrency Price}}\\
Shalini Sharma$^{a}$ (shalinis@iiitd.ac.in), Angshul Majumdar$^b$ (angshul@iiitd.ac.in)\cred{, Emilie Chouzenoux$^c$ (emilie.chouzenoux@inria.fr), V\'{i}ctor Elvira$^d$ (victor.elvira@ed.ac.uk)}\\

\hspace{10pt}

\begin{flushleft}
\small  
$^a$ {PhD Scholar, Indraprastha Institute of Information Technology-Delhi, India} \\
$^b$ Associate Professor, Indraprastha Institute of Information Technology-Delhi, India \\
\cred{
$^c$ Research Director, Inria Saclay, University Paris Saclay, France\\
$^d$ Professor, School of Mathematics, University of Edinburgh, UK
}

\vspace{1cm}
\textbf{Corresponding Author:} \\
Shalini Sharma\\
PhD Scholar, Indraprastha Institute of Information Technology-Delhi, India \\
Email: shalinis@iiitd.ac.in.

\end{flushleft}        
\end{center}
\end{titlepage}

\title{\cmazenta{A Deep State-Space Model for Predicting Cryptocurrency Price}}

\author[1]{Shalini Sharma}
\ead{shalinis@iiitd.ac.in}

\author[2]{Angshul Majumdar}
\ead{angshul@iiitd.ac.in}

\author[3]{Emilie Chouzenoux}
\ead{emilie.chouzenoux@inria.fr}

\author[4]{Victor Elvira}
\ead{victor.elvira@ed.ac.uk}

\cortext[cor1]{Corresponding author.}
\address[1]{PhD Scholar, Indraprastha Institute of Information Technology-Delhi, India}
\address[2]{Associate Professor, Indraprastha Institute of Information Technology-Delhi, India}
\address[3]{Research Director, Inria Saclay, University Paris Saclay, France}
\address[4]{Professor, School of Mathematics, University of Edinburgh, UK}

\begin{abstract}

\cred{Our work presents two fundamental contributions. On the application side, we tackle the challenging problem of predicting day-ahead crypto-currency prices. On the methodological side, a new dynamical modeling approach is proposed. Our approach keeps the probabilistic formulation of the state-space model, which provides uncertainty quantification on the estimates, and the function approximation ability of deep neural networks. We call the proposed approach the deep state-space model. The experiments are carried out on established cryptocurrencies (obtained from Yahoo Finance). The goal of the work has been to predict the price for the next day. Benchmarking has been done with both state-of-the-art and classical dynamical modeling techniques. Results show that the proposed approach yields the best overall results in terms of accuracy.} 
\end{abstract}


\end{frontmatter}

\section{Introduction}
\label{introduction}

Investopedia defines crypto-currency as “a digital or virtual currency that is secured by cryptography, which makes it nearly impossible to counterfeit or double-spend” and is built on “decentralized networks based on block-chain technology—a distributed ledger enforced by a disparate network of computers”. A defining feature of crypto-currencies is that they are usually not issued by central banking agencies like the Federal Reserve System in US, Bank of Canada, European Central Bank, or the People’s Bank of China; this makes crypto-currencies (theoretically) immune to government interventions. 

The introduction of Bitcoin around 2009 and its meteoric rise led to investors infuse their funds in crypto-currencies. One major reason behind the shift in investment largely owes to the 2008 financial crisis, which subsequently led to the waning of trust in the banking system. The market capitalization of crypto-currencies rose from less than 10 billion in 2014 to more than 2 trillion in 2021.

However, crypto-currencies are extremely volatile. To give an example, the volatility index of the most stable crypto-currency \cblue{USD Tether (USDT)} has been between 95 and 100 in August 2021, while that of a blue chip corporation like \cblue{Microsoft (MSFT)} has been around 16 in the same period; a highly volatile smallcap stock like \cblue{Genworth Financial (GNW)} in the same period had a volatility index less than 40. Such large volatility makes predicting crypto-currency prices a more challenging problem than stock forecasting. The reason crypto-currencies are volatile is because they do not have any intrinsic value. Their prices are mainly dependent on the emotion of investors, and in such a scenario, tweets from influencers can play a major role in swaying their prices; one example of how tweets from a major influencer can drive prices high or low can be seen from \citep{molla2021elon,Financial_express}.

\cmazenta{This work addresses the most challenging problem in personal finance today – forecasting - prices. There are a few studies on this subject. Recent study \citep{livieris2021advanced},\citep{ye2022relationship} use off-the-shelf deep learning tools for predicting crypto-currency prices. Competitive survey analysis of forecasting crypto-currency using machine learning method can be found \citep{derbentsev2020forecasting}. A different branch of study \citep{yasir2020deep} follows cues from social media for predicting the crypto-currency and gives more insights for estimating future prices \citep{kraaijeveld2020predictive}. One of the study uses ARCH-MIDAS framework to identify drivers of Crypto-currency volatility \citep{walther2019exogenous}. The recent study uses the predictive power of social signals, specifically user behavior and communication patterns for forecasting prices for cypto-currencies \citep{glenski2019improved}.} The main shortcoming of all the said studies is that they yield point predictions; given the volatility of the crypto-currency market, the investor needs to know price value and the prediction uncertainty. None of the prior studies can provide that. This is the reason predicting the volatility of cryptocurrencies is a separate branch of research \citep{ma2020cryptocurrency,catania2018predicting, kristjanpoller2018hybrid, kochling2020volatility}. 

Current research in cryptocurrency forecasting offers a piecemeal solution – one approach for predicting price and a separate one for predicting volatility. This is as good as comparing apples to oranges since the fundamental models and assumptions of the two approaches will be different. \cmazenta{The paper} propose a single model that yields both the point estimate of cryptocurrency prices as well as the uncertainty around the estimate. 

\cred{\cmazenta{The work} is based on the classical \cblue{state-space model (SSM)} for time-series analysis. SSM is defined by two functions: the   Markovian state model and the observation model. In the traditional approach, the models are assumed to be known, but in realistic financial forecasting applications this is never the case.} This is the reason prior studies \citep{sharma2021recurrent,sharma2021sequential} proposed to learn the models instead; the two aforementioned papers are similar in principle and only vary in the introduction of an exogenous input in the model. However, these works were based on linear state-evolution and observation models and hence could only model piece-wise linear functions. \cmazenta{In this work, the above said restriction is removed by proposing to model} the underlying functions by deep learning. This results in the so-called deep SSM. The underlying SSM structure ensures that we can both predict a point estimate as well quantify the uncertainty about it, making it a perfect fit for cryptocurrency forecasting.

\cblue{\cmazenta{The proposed work} introduces deep non-negative matrix factorization (deep NMF) models for learning the approximations on operators. Deep NMF \citep{de2021survey} is equivalent to Deep Rectified Linear Unit (ReLU) networks. There is existing literature \citep{tariyal2016deep,mahdizadehaghdam2019deep} which establishes its connection with deep dictionary learning; the first paper is a more generalised non-linear version of the later. \cmazenta{The work} have utilised the potential of deep NMF / ReLU in \cmazenta{the proposed work} by embedding it into a Gaussian SSM. This allows for modelling non-linearity of the underlying dynamical process. The main difference between the prior shallow model and the proposed deep one is that the \cmazenta{proposed work} is more generalized version of the former. The shallow model can only approximate piece-wise linear functions; the proposed one, being formulated on ReLU network can approximate arbitrary non-linear and non-smooth functions. The price we pay for the generalisation ability of our approach is the difficulty in training. The shallow model allowed for closed form solutions; the proposed deeper extension does not. Hence, \cmazenta{the work} resort to \cred{alternating majorization-minimization (AMM)} approach to solve this. \cmazenta{The proposed approach is named as Deep state-space model for Predicting Cryptocurrency Price (DeCrypt). }}

\cblue{The rest of the paper is organized as follows. Related work in  literature is reviewed in Section 2. The proposed model and inference algorithm are explained in Section 3. The application relevance is discussed in Section 4. The experimental results are presented and discussed in Section 5. The acknowledgement is discussed in Section 6.
Conclusions and future directions of research are finally given in Section 7.}

\section{Related Work}
The objective of this work is to predict cryptocurrency prices. Although the problem is new, it is akin to the problem of stock forecasting in particular and financial forecasting in general. \cred{There are two approaches to address such time-series modelling problems. The first approach comprises of state-space-model (SSM) and auto-regressive moving average (ARMA). These methods are usually employed in signal processing and statistical applications (e.g., in statistical ecology \cite{newman2023state}). The second approach is based on machine learning methods, more specifically on recurrent neural network (RNN)}. 

Signal processing techniques are interpretable and yield uncertainty estimates. However, the problem with SSM and ARMA is that their model parameters need to be known. The pros and cons of such assumptions have been studied for the linear \citep{kalmanthesis} and non-linear \citep{andersen2009handbook} cases. Specification of the underlying models lead to simplistic (and restrictive) models that fail to capture the movement of stock prices. ARMA \citep{rounaghi2016investigation} and it variants like \cmazenta{Autoregressive Integrated Moving Average (ARIMA)} \citep{jarrett2011arima} (Box Jenkins models in general \citep{dritsaki2015box}) have been widely used in financial forecasting. These too require specification of underlying parameters defining the price movement; a higher order model fits the training data but fails to generalize and a lower order model yields poor results both on training and testing data. 

RNN on the other hand can learn the underlying function from the training data; thanks to their function approximation ability \citep{BarbaraHammer, garzon1999dynamical}. They do not need specifying any function parameter, given enough data they can learn the underlying dynamical model. This is the reason they have been more successful in recent years for financial forecasting \citep{baek2018modaugnet, kim2019forecasting}. The shortcoming of RNN is that in their vanilla form, they do not yield uncertainty estimates – we have already discussed its importance in the introduction. This is the reason, researchers are concentrating on building RNN models on probabilistic frameworks \citep{rangapuram2018deep,ma2020particle}. 

One must note the fundamental difference between our proposal and previous approaches like \citep{rangapuram2018deep,ma2020particle}. In \citep{rangapuram2018deep} the state-evolution of equation of SSM is modelled as a recurrent neural network, but the observation is assumed to be known with further restrictions of linearity and incoherence. Our model embeds deep neural networks in both the state-evolution and observation models, without any restrictions. One can assume that ours is a more generalized version of \citep{rangapuram2018deep}. The work \citep{ma2020particle} is in some sense complimentary to \citep{rangapuram2018deep}; they embed a particle filter in an RNN thereby adding a probabilistic flavour to the otherwise deterministic latent states (see more on particle filters in \citep{sarkka2013bayesian,elvira2017adapting} for more details). 

Several prior studies such as \citep{digalakis1993ml,sharma2020blind} have proposed solutions for the linear SSM when both the state-evolution and observation matrices are unknown. These techniques were able to learn piece-wise linear functions from the data; however they were not able to model arbitrary functions. The work overcomes this limitation by modelling the state-evolution and observation matrices as deep neural networks. Recent papers such as \citep{liang2016deep,elbrachter2021deep} are showing how deep neural networks excel over shallow networks in terms of function approximation; the work is rooted on the same. 
\section{Proposed Method}
\cmazenta{This section will discuss the proposed approach (Deep State-Space Model for Predicting Cryptocurrency Price (DeCrypt)) in detail. For convenience the proposed method will be referred by name \textbf{DeCrypt}.} 
\subsection{\cmazenta{Model Details}}
\label{sec:model}
\cmazenta{The proposed work} is based on standard state-space model (SSM). It can be expressed as : \\
For every $k \in \{1,\ldots,K\}$:
 \begin{equation}
  \left\{ 
  \begin{array}{ll}
      \z_{k} &= f(\z_{k-1})+ g(\u_{k}) + \v_{1,k},   \\
      \x_k &= h(\z_k) + \v_{2,k}.
    \end{array}
    \right.
    \label{eq:func}
    \end{equation}
\cblue{The goal is to infer $(\z_k)_{1 \leq k \leq K}$, a sequence of unknown latent space vector of size $N_z \geq 1$ given the  input $(\u_k)_{1 \leq k \leq K}$ vector of size $N_y\geq 1$ and observed sequence $(\x_k)_{1 \leq k \leq K}$ of vector of size $N_x \geq 1$. \cmazenta{The work} assumes process noises $(\v_{1,k})_{1 \leq k \leq K}$, $(\v_{2,k})_{1 \leq k \leq K}$ to have a Gaussian distribution with zero-mean and covariance matrix $\mathbf{Q}$ and $\mathbf{R}$, respectively. The covariance matrices are symmetric definite positive. Here $K$ is the total number of data to be processed (window size in our case).}

Traditional solutions to SSM required the functions $f(.), g(.)$ and $h(.)$ to be known. When the functions are linear, prior studies \citep{digalakis1993ml, sharma2020blind} proposed a solution called blind Kalman filtering; blind since the functions/matrices were assumed to be unknown. \cred{Recent extensions introduced a graphical perspective for $f$ (still assumed to be linear), along with suitable sparse priors \cite{ElviraGraphEM2022,cox2023sparse,chouzenoux2023graphit}.}

\cmazenta{This work} removes the linearity restriction, by \cblue{embedding ReLU deep neural networks (DNNs) }in place of the functions. Our model thus takes the form:\\
For every $k \in \{1,\ldots,K\}$:
 \begin{equation}
  \left\{ 
  \begin{array}{ll}
      \z_{k} &= \T_{10}\T_{11}\T_{12} \z_{k-1} + \T_{20}\T_{21}\T_{22} \u_{k} + \v_{1,k},   \\
      \x_k &= \D_0\D_1\D_2 \z_k + \v_{2,k}.
    \end{array}
    \right.
    \label{eq:model}
    \end{equation}
This is a multi-linear Gaussian model; everything except the input $(\u_k)_{1 \leq k \leq K}$ and  observed  sequence $(\x_k)_{1 \leq k \leq K}$ are unknown. \cblue{The primary objective is to jointly learn the latent factor matrices $\T_{10} \in \mathbb{R}^{N_z \times N_z}$, $\T_{11} \in \mathbb{R}^{N_z \times N_z}$, $\T_{12} \in \mathbb{R}^{N_z \times N_z}$, three positive-valued linear factors leading to a multi-linear state operator $\T_{10}\T_{11}\T_{12}$, the control input transition matrices $\T_{20} \in \mathbb{R}^{N_z \times N_z}$, $\T_{21} \in \mathbb{R}^{N_z \times N_z}$, $\T_{22} \in \mathbb{R}^{N_z \times N_y}$, three positive-valued linear factors leading to a multi-linear control operator $\T_{20}\T_{21}\T_{22}$, and the observation matrices $\D_0 \in \mathbb{R}^{N_x \times N_z}$, $\D_1 \in \mathbb{R}^{N_z \times N_z}$, $\D_2\in \mathbb{R}^{N_z \times N_z}$, three positive-valued linear factors yielding the multi-linear observation model $\D_{0}\D_{1}\D_{2}$ and the sequence $(\z_k)_{1 \leq k \leq K}$, from observed sequence $(\x_k)_{1 \leq k \leq K}$ and $(\u_k)_{1 \leq k \leq K}$.} \footnote{Throughout the paper, three-terms factorizations is considered , for the sake of readability. \cblue{The 3-layers modeling and inference methodology has the great advantage of being generic enough to be straightforwardly extended to any number, greater or equals to one, of factors.}}
The inference problem can be categorised as a blind filtering problem, where the objective is to infer the time series predictions and unknown model parameters from the given input data and observed sequences. \cblue{As stated earlier, the classical SSM techniques need prior assumptions and information on model parameters. In the proposed model described here, this would imply explicitly setting some prior values to the positive latent factor matrices matrices $\{\T_{10},\T_{11},\T_{12},\T_{20},\T_{21},\T_{22},\D_0,\D_1,\D_2\}$ involved in both state, control and observation models. In real-world applications, especially in areas as complex as financial modelling this is never known; this is largely owing to the non-stationarity and volatility of the process. The main objective of the proposed work is to provide a point-wise estimate of the positive latent factor matrices $\{\T_{10},\T_{11},\T_{12},\T_{20},\T_{21},\T_{22},\D_0,\D_1,\D_2\}$ and obtain a probabilistic estimate of sequence $(\z_k)_{1 \leq k \leq K}$, given the observed sequence $(\x_k)_{1 \leq k \leq K}$ and control input $(\u_k)_{1 \leq k \leq K}$. Therefore \cmazenta{the work} propose to jointly solve for (i) the three deep NMF problems, and (ii) the filtering/smoothing problem.}

\subsection{\cblue{Model Analysis}}
\begin{figure}[!ht]
    \centering
    \includegraphics[width=11cm]{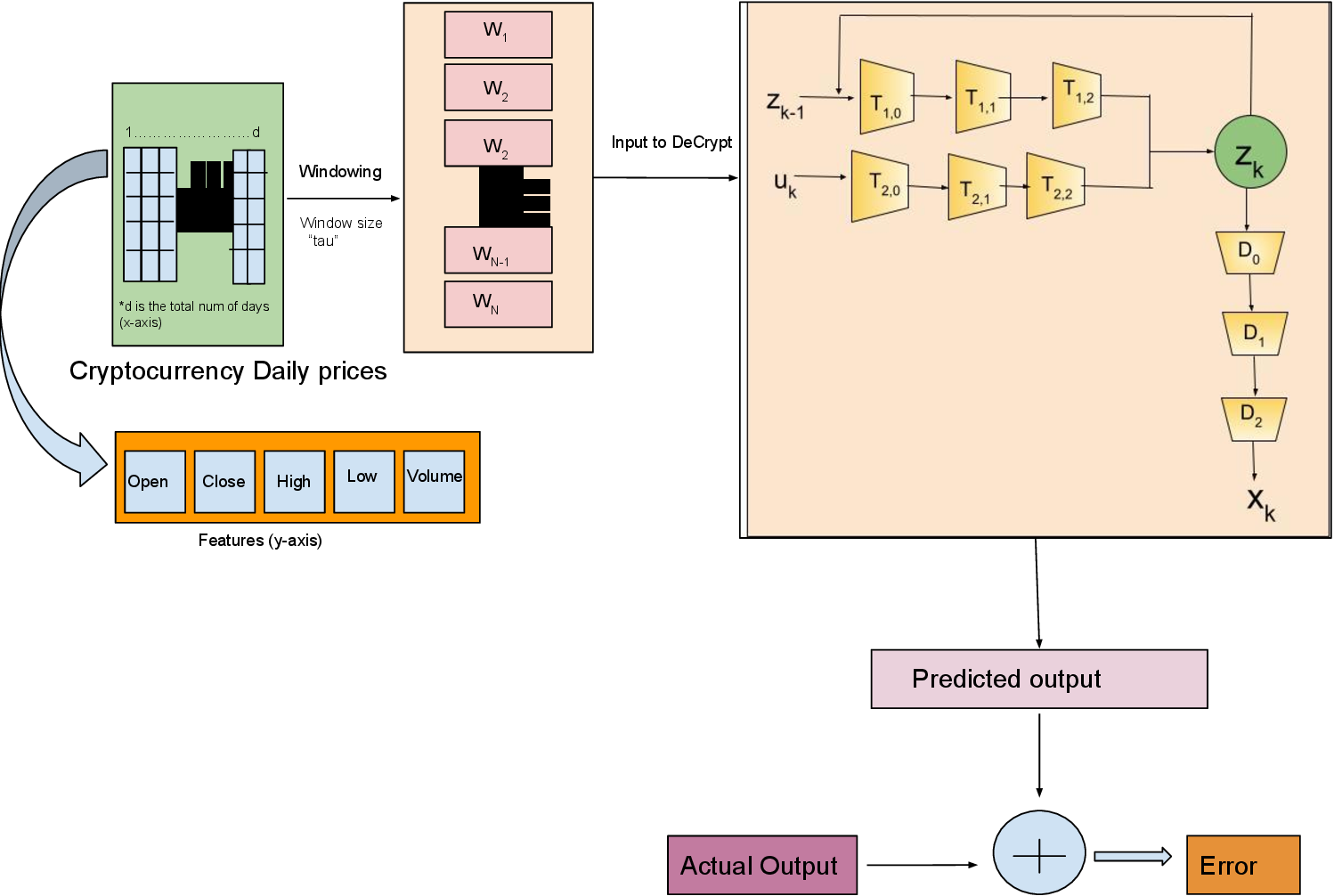}
    \caption{Schematic Diagram for Proposed model DeCrypt.}
    \label{fig:schematic}
\end{figure}

\cblue{This section will describe the fundamental characteristics of the proposed DeCrypt approach. 
We have used the discrete time invariant state-space model with exogenous input. The mathematical fundamentals of the model in Eq. \eqref{eq:model} and schematic diagram in Fig ~\ref{fig:schematic}. The former part of the equation works on the evolution of the hidden state parameters, where the model assumes Markovianity between two consecutive hidden states. The later part defines the relationship between the hidden and observed states. We depart from all prior works in SSM based dynamical modelling in the way we define the functions. Usually a matrix is used when the functions are assumed to be non-linear and an explicit non-linear function otherwise \citep{andrieu2010particle,chopin2013smc2,crisan2018nested}. Here we model non-linearity by a deep ReLU network; alternately this can be also seen as a  deep NMF \citep{de2021survey}. The reason for using a deep ReLU network is its universal function approximation capbility~\citep{liu2021optimal,daubechies2022nonlinear,chen2019efficient}. It is essential to note here that classical state-space models uses Monte Carlo simulation or Variable Bayes type techniques, doe non-linear SSM. These techniques are complex and usually do not scale well. In contrast, in the proposed method, each layer is modeled and learnt in the form of matrix that can be estimated using an \cred{alternating majorization-minimization (AMM) procedure}.}

\cblue{When compared to existing literature in machine learning approaches, DNN mostly utilizes in backpropagation for its training \citep{Chen2021deepNMF,flenner2017deep}. Consequently while modelling dynamical systems, \cmazenta{backpropagagtion through time (BPTT)} needs to be employed. We are all aware of the pitfalls of  BPTT. This is the reason we resort to \cred{AMM} instead. Unlike BPTT, \cred{AMM} (under certain conditions) at least guarantees convergence to a local minimum~\cite{Chouzenoux2016,jacobson}. }
\subsection{\cmazenta{Model Inference Algorithm}}

\cblue{The inference problem can be viewed as smoothing/filtering problem where we aim to infer probabilistic estimate of the hidden state $(\z_k)_{1 \leq k \leq K}$. In this work, we have also introduced deep NMF factors (described earlier) which are unknown. We aim to jointly infer both probabilistic distribution on hidden state and deep NMF factors estimation from the data.} To estimate the state matrices, control input transition matrices and observation transition matrices, we use Expectation-maximization strategy (for more details \citep[chap.12]{sarkka2013bayesian} and \citep{shumway1982approach}. The EM strategy operates in two steps namely E-step where we assume the \cblue{positive} latent factor matrices $\{\T_{10},\T_{11},\T_{12},\T_{20},\T_{21},\T_{22},\D_0,\D_1,\D_2\}$ to be fixed and estimate probabilistic inference for the state representation $(\z_k)_{1 \leq k \leq K}$. M-step involves updating these matrices assuming fixed state (learnt from E-step). The E-step is akin to that of a Kalman filter / smoother. The M-step updates the matrices $\{\T_{10},\T_{11},\T_{12},\T_{20},\T_{21},\T_{22},\D_0,\D_1,\D_2\}$ by maximizing the upper bound : 
\begin{multline}
    \varphi_K(\T_{10},\T_{11},\T_{12},\T_{20},\T_{21},\T_{22},\D_0,\D_1,\D_2)\\ = \log p(\x_{1:K}| \T_{10},\T_{11},\T_{12},\T_{20},\T_{21},\T_{22},\D_0,\D_1,\D_2).
\end{multline}
It is important to note that the inference of the $i+1$-th EM update is obtained from the estimates from the previous iteration $i$. We explain the EM algorithm in more detail. 
\subsubsection{E-step: Kalman/RTS inference}
\label{sec:kalman}
We consider the latent factors $\T_{10}^{[i]}$, $\T_{11}^{[i]}$, $\T_{12}^{[i]}$, $\T_{20}^{[i]}$, $\T_{21}^{[i]}$, $\T_{22}^{[i]}$, $\D_0^{[i]}$, $\D_1^{[i]}$, $\D_2^{[i]}$ to be fixed. The objective of this step is to infer the probabilistic estimation of the state. The initial state takes the form $\z_0 \sim \mathcal{N}(\bar{\z}_0, \P_0)$ with $\bar{\z}_0 \in \mathbb{R}$ and $\P_0$ defined as definite symmetric positive matrix $\in$ $\mathbb{R}^{N_z \times N_z}$. 
The probabilistic estimation is provided by the Kalman filter through predictive distribution :
\begin{equation}
    p(\z_k|\x_{1:k},\u_{1:k}) = \mathcal{N}(\z_k; \bar{\z}_k,\P_k).
    \label{Kal_output}
\end{equation}
For every $k$, the mean $\bar{\z}_k$ and the covariance $\P_k$ are given by the Kalman iterations:\\
\noindent For $k=1,\ldots,K$:\\
\textit{Predict state:}
\small
\begin{equation}
            \left\{ 
    \begin{array}{ll}
            \z_k^-& = \T_{10}^{[i]}\T_{11}^{[i]}\T_{12}^{[i]}\bar \z_{k-1}+ \T_{20}^{[i]}\T_{21}^{[i]}\T_{22}^{[i]} \u_{k},   \\
        \P_k^-& =\T_{10}^{[i]}\T_{11}^{[i]}\T_{12}^{[i]}\P_{k-1}(\T_{10}^{[i]}\T_{11}^{[i]}\T_{12}^{[i]})^\top + \Q.
        \end{array}
        \right.
        \label{eq:predict}
        \end{equation} 

\textit{Update state:}   
\small
    \begin{equation}
            \left\{ 
    \begin{array}{ll}
            \y_k&= \x_k-\D_0^{[i]}\D_1^{[i]}\D_2^{[i]}\z_k^-,   \\
            \S_k&=\D_0^{[i]}\D_1^{[i]}\D_2^{[i]}\P_k^-(\D_0^{i}\D_1^{[i]}\D_2^{[i]})^\top +\R,  \\
            \K_k&=\P_k^-(\D_0^{[i]}\D_1^{[i]}\D_2^{[i]})^\top \S_k^{-1}, \\
            \z_k&=\z_k^- +\K_k \y_k,  \\
            \P_k&=\P_k^--\K_k\S_k\K_k^\top.
        \end{array}
        \right.
        \label{update_kalman}
        \end{equation}

\normalsize
Hereabove, $\y_k$ represents the measurement pre-fit residual, $\S_k$ represents the pre-fit covariance, $\K_k$ represents Kalman gain, $\bar \z_k$ represents the updated (a posteriori) state estimate, $\P_k$ represents the updated (a posteriori)  covariance estimate. The backward recursion from the RTS smoother allow to build the smoothing distribution $p(\z_k | \x_{1:K}, \u_{1:K})$.
\noindent For $k=K, \ldots, 1$ \\
\textit{Backward Recursion (Bayesian Smoothing):} 
\small   
     \begin{equation}
            \left\{ 
    \begin{array}{ll}
     
            \z_{k+1}^-&= \T_{10}^{[i]}\T_{11}^{[i]}\T_{12}^{[i]}\bar \z_k+ \T_{20}^{[i]}\T_{21}^{[i]}\T_{22}^{[i]} \u_{k},   \\
            \P_{k+1}^-&=\T_{10}^{[i]}\T_{11}^{[i]}\T_{12}^{[i]}\P_k(\T_{10}^{[i]}\T_{11}^{[i]}\T_{12}^{[i]})^\top +\Q,  \\
            \G_k&=\P_k(\T_{10}^{[i]}\T_{11}^{[i]}\T_{12}^{[i]})^\top [\P_{k+1}^{-}]^{-1},      \\
            \z_k^s&=\z_k +\G_k[\z_{k+1}^s-\z_{k+1}^-],  \\
            \P_k^s&=\P_k+\G_k[\P_{k+1}^s-\P_{k+1}^-]\G_k^\top.
        \end{array}
        \right.
        \label{eq:smooth}
        \end{equation}
\normalsize
Consequently, for every time step $k \in \{1,\ldots,K\}$, the RTS smoother  provides:
\begin{equation}
 {p(\z_k | \x_{1:K},\u_{1:K}) = \mathcal{N}(\z_k;\z_k^s,\P_k^s)}. \label{eq:RTSoutput}
\end{equation}
\subsubsection{M-step: Operator update}
This step utilizes the estimated state ($\z_k$) following an optimization step to increase the likelihood of the matrix parameters $\T_{10}$, $\T_{11}$, $\T_{12}$, $\T_{20}$, $\T_{21}$, $\T_{22}$, $\D_0$, $\D_1$, $\D_2$, using the smoothed predictive distribution obtained in the E-step. 

\begin{multline}
    \varphi_K(\T_{10},\T_{11},\T_{12},\T_{20},\T_{21},\T_{22},\D_0,\D_1,\D_2)\\ \geq \mathcal{Q}( \T_{10},\T_{11},\T_{12},\T_{20},\T_{21},\T_{22},\D_0,\D_1,\D_2;\mathbf{\Theta}^{[i]}).
\end{multline} 
Here above,  $\mathbf{\Theta}^{[i]} = \{\Sigmab^{[i]},\Phib^{[i]},\B^{[i]},\C^{[i]},\Deltab^{[i]},\A^{[i]},\F^{[i]},\I^{[i]}\}$ gathers eight quantities \cblue{(variables)} defined from the outputs of the  E-step described in Sec.~\ref{sec:kalman}):
\begin{multline}
 \mathcal{\Q}(\T_{10},\T_{11},\T_{12},\T_{20},\T_{21},\T_{22},\D_0,\D_1,\D_2; \mathbf{\Theta}^{[i]}) = \\
+ \frac{K}{2}  \text{tr} \left( \Q^{-1} (\Sigmab^{[i]} - (\T_{10}\T_{11}\T_{12})^\top \C^{[i]}-\A^{[i]} (\T_{20}\T_{21}\T_{22})^\top -(\T_{10}\T_{11}\T_{12})(\C^{[i]})^\top \right. \\ \left.
+(\T_{10}\T_{11}\T_{12})\Phib^{[i]} (\T_{10}\T_{11}\T_{12})^\top +(\T_{10}\T_{11}\T_{12}) \F^{[i]} (\T_{20}\T_{21}\T_{22})^\top - (\T_{20}\T_{21}\T_{22})(\A^{[i]})^\top \right.\\  
\left.+(\T_{20}\T_{21}\T_{22})(\F^{[i]})^\top (\T_{10}\T_{11}\T_{12})^\top
+ (\T_{20}\T_{21}\T_{22}) \I^{[i]} (\T_{20}\T_{21}\T_{22})^\top) \right) \\
+ \frac{K}{2}  \text{tr} \left(\R^{-1} \mathbf{\Deltab}^{[i]} - \B^{[i]} (\D_0 \D_1  \D_2)^\top - \D_0 \D_1 \D_2  (\B^{[i]})^\top 
 + \D_0 \D_1  \D_2  \Sigmab^{[i]} (\D_0 \D_1  \D_2 )^\top \right),
\end{multline}
with: 
\begin{equation}
\small
            \left\{ 
    \begin{array}{ll}
\Sigmab ^{[i]}& = \frac{1}{K} \sum_{k=1}^K \P_k^s + \z_k^s (\z_k^s)^\top, \\  
    \Phib^{[i]} & = \frac{1}{K} \sum_{k=1}^K \P_{k-1}^s + \z_{k-1}^s (\z_{k-1}^s)^\top,\\  
     \B^{[i]} &= \frac{1}{K} \sum_{k=1}^K \x_k (\z_k^s)^\top, \\  
     \C^{[i]} &= \frac{1}{K} \sum_{k=1}^K (\P_k^s \G_{k-1}^\top + \z_k^s(\z_{k-1}^s)^\top).\\
     \label{eq:RTSquantities}
     \A^{[i]} &=\frac{1}{K}\sum_{k=1}^{K} \z^{s}_k \u^\top_{k} \\
     \F^{[i]} &=\frac{1}{K}\sum_{k=1}^{K} \z^{s}_{k-1} \u^\top_{k} \\
     \I^{[i]} &=\frac{1}{K}\sum_{k=1}^{K} \u_k \u^\top_{k} \\
     \Deltab^{[i]} &=\frac{1}{K}\sum_{k=1}^{K} \x_k \x_{k}^\top
     
\end{array}
\right.
\end{equation}
\normalsize
In this step lies the main bottleneck of the deeper extension (compared to the shallow state-space model of \citep{sharma2021sequential}). For the shallow model each of the operators was a single matrix; therefore there update step resulted in a linear inverse problem. Such is not the current case; here the variables are multi-linear in nature. Therefore we do not have the simple (analytic) updates - as was the case for the shallow model. We have to resort to the paradigm of alternating direction method of multipliers (ADMM)\citep{wang2019global,nishihara2015general,lin2015global}
for solving updating the variables from the multi-linear form. In ADMM, the idea is that, one can update one variable assuming the others to be constant and as long as each of the variables have a closed form update, the overall optimization will reach a local minimum. Based on the ADMM approach the computations each variable \cblue{under the positivity constraints on the factors} $\T_{10}^{[i+1]},\T_{11}^{[i+1]},\T_{12}^{[i+1]}$, $\T_{20}^{[i+1]},\T_{21}^{[i+1]},\T_{22}^{[i+1]}$ and $\D_0^{[i+1]}$, $\D_1^{[i+1]}$, $\D_2^{[i+1]}$ leads to: 

\small{
\begin{align*}
(\T_{10})^{[i+1]} = \underset{\T_{10}\cblue{\geq 0}}{\rm{argmax}}  \mathcal{\Q}(\T_{10},\T_{11}^{[i]},\T_{12}^{[i]},\T_{20}^{[i]},\T_{21}^{[i]},\T_{22}^{[i]}\D_0^{[i]},\D_1^{[i]},\D_2^{[i]}; \mathbf{\Theta}^{[i]})\\
(\T_{11})^{[i+1]}  = \underset{\T_{11}\cblue{\geq 0}}{\rm{argmax}}  \mathcal{\Q}(\T_{10}^{[i+1]},\T_{11},\T_{12}^{[i]},\T_{20}^{[i]},\T_{21}^{[i]},\T_{22}^{[i]} D_0^{[i]},\D_1^{[i]},\D_2^{[i]}; \mathbf{\Theta}^{[i]}) \\
(\T_{12})^{[i+1]} = \underset{\T_{12}\cblue{\geq 0}}{\rm{argmax}}  \mathcal{\Q}(\T_{10}^{[i+1]},\T_{11}^{[i+1]},\T_{12},\T_{20}^{[i]},\T_{21}^{[i]},\T_{22}^{[i]}\D_0^{[i]},\D_1^{[i]},\D_2^{[i]}; \mathbf{\Theta}^{[i]}) \\
(\T_{20})^{[i+1]}  = \underset{\T_{20}\cblue{\geq 0}}{\rm{argmax}}  \mathcal{\Q}(\T_{10}^{[i+1]},\T_{11}^{[i+1]},\T_{12}^{[i+1]},\T_{20},\T_{21}^{[i]},\T_{22}^{[i]} \D_0^{[i]},\D_1^{[i]},\D_2^{[i]}; \mathbf{\Theta}^{[i]})\\
(\T_{21})^{[i+1]}  = \underset{\T_{21}\cblue{\geq 0}}{\rm{argmax}}  \mathcal{\Q}(\T_{10}^{[i+1]},\T_{11}^{[i+1]},\T_{12}^{[i+1]},\T_{20}^{[i+1]},\T_{21},\T_{22}^{[i]}\D_0^{[i]},\D_1^{[i]},\D_2^{[i]}; \mathbf{\Theta}^{[i]})\\
(\T_{22})^{[i+1]}  = \underset{\T_{22}\cblue{\geq 0}}{\rm{argmax}}  \mathcal{\Q}(\T_{10}^{[i+1]},\T_{11}^{[i+1]},\T_{12}^{[i+1]},\T_{20}^{[i+1]},\T_{21}^{[i+1]},\T_{22}\D_0^{[i]},\D_1^{[i]},\D_2^{[i]}; \mathbf{\Theta}^{[i]})\\
(\D_{0})^{[i+1]}  = \underset{\D_{0}\cblue{\geq 0}}{\rm{argmax}}  \mathcal{\Q}(\T_{10}^{[i+1]},\T_{11}^{[i+1]},\T_{12}^{[i+1]},\T_{20},\T_{21}^{[i+1]},\T_{22}^{[i+1]}\D_0,\D_1^{[i]},\D_2^{[i]}; \mathbf{\Theta}^{[i]})\\
(\D_{1})^{[i+1]}  = \underset{\D_{1}\cblue{\geq 0}}{\rm{argmax}}  \mathcal{\Q}(\T_{10}^{[i+1]},\T_{11}^{[i+1]},\T_{12}^{[i+1]},\T_{20}^{[i+1]},\T_{21}^{[i+1]},\T_{22}^{[i]},\D_0^{[i]},\D_1,\D_2^{[i]}; \mathbf{\Theta}^{[i]})\\
(\D_{2})^{[i+1]}  = \underset{\D_{2}\cblue{\geq 0}}{\rm{argmax}}  \mathcal{\Q}(\T_{10}^{[i+1]},\T_{11}^{[i+1]},\T_{12}^{[i+1]},\T_{20}^{[i+1]},\T_{21}^{[i+1]},\T_{22}^{[i+1]}\D_0^{[i+1]},\D_1^{[i+1]},\D_2; \mathbf{\Theta}^{[i]})
\end{align*}}
\normalsize
The above sub-problems can easily be rewritten as the minimization of convex quadratic functions which can be solved through several solvers. We stick to use simple \cred{projected least-squares updates}, which is also reminiscent from the literature of deep nonnegative matrix factorization \citep{Chen2021deepNMF}, and the deep ReLu neural networks models \citep{daubechies2022nonlinear}. The deep neural network (DNN) based operators are regularized by imposing a positivity constraint on the entries of the estimated matrices, by simply projecting them onto the positive orthant after each update of the M-step (similar to ReLU activation function). This is akin to deep non-negative matrix factorization  \citep{trigeorgis2016deep,Mei}, while keeping the convergence behaviour of the EM algorithm. DNN with ReLU activation is known for its function approximation ability \citep{chen2019efficient,yarotsky2018optimal}. This yields the following analytic updates:\\
\cblue{
\small{
\begin{align}
\T_{10}^{[i+1]} &=\text{ReLu} \left(\bigl(\C^{[i]}(\T_{12}^{[i]})^\top(\T_{11}^{[i]})^\top \bigr)\notag-\bigl (\T_{20}^{[i]}\T_{21}^{[i]}\T_{22}^{[i]}(\F^{[i]})^\top(\T_{12}^{[i]})^\top(\T_{11}^{[i]})^\top \bigr)\notag \right.\\
& \left. \qquad \times (\T_{11}^{[i]}\T_{12}^{[i]}\Phib^{[i]}(\T_{12}^{[i]})^\top(\T_{11}^{[i]})^\top))^{\dagger}\right)\notag\\
\T_{11}^{[i+1]}&=\text{ReLu}\left(((\T_{10}^{[i+1]})^\top\Q^{-1}(\T_{10}^{[i+1]})^{-1})\bigl(\bigl((\T_{10}^{i+1})^\top\Q^{-1}\C^{[i]}\T_{12}^{[i]}\bigr)\right.\notag \\
 & \left. \qquad -\bigl(\T_{10}^{[i+1]}\Q^{-1}\T_{20}^{[i]}\T_{21}^{[i]}\T_{22}^{[i]}(\F^{[i]})^\top\T_{12}^{[i]}\bigr)\bigr)\notag
\times (\T_{12}^{[i]}\Phib^{[i]}(\T_{12}^{[i]})^\top)^{\dagger}\right) \notag\\
\T_{12}^{[i+1]}&=\text{ReLu}\left(((\T_{11}^{[i+1]})^\top(\T_{10}^{[i+1]})^\top\Q^{-1}\T_{10}^{[i+1]}\T_{11}^{[i+1]})^{\dagger} \notag
 \times \bigl((\T_{11}^{[i+1]})^\top(\T_{10}^{[i+1]})^\top\C^{[i]}\Q^{-1}\bigr)\notag \right.\\
 & \left. \qquad - \bigl((\T_{11}^{[i+1]})^\top(\T_{10}^{[i+1]})^\top(\Q^{-1}\T_{20}^{[i]}\T_{21}^{[i]}\T_{22}^{[i]}(\F^{[i]})^\top)\Phib^{-1}\bigr)\right)\notag\\
\T_{20}^{[i+1]}&=\text{ReLu}\left(\bigl((\T_{22}^{[i]})^\top(\T_{21}^{[i]})^\top-\T_{10}^{[i+1]}\T_{11}^{[i+1]}\T_{12}^{[i+1]}\F^{[i]}(\T_{22}^{[i]})^\top(\T_{21}^{[i]})^\top\bigr)\notag \right. \\
& \left. \qquad \times(\T_{21}^{[i]}\T_{22}^{[i]}\I^{[i]}(\T_{21}^{[i]})^\top(\T_{22}^{[i]})^\top))^{\dagger}\right)\notag\\
\T_{21}^{[i+1]}&=\text{ReLu}\left(\bigl(\T_{20}^{[i+1]}\Q^{-1}(\T_{20}^{[i+1]})^\top \bigr)^{\dagger} \bigl(\T_{20}^{[i+1]}\A^{[i+1]}\Q^{-1}(\T_{22}^{[i]})^\top \notag \right.\\
& \qquad - \bigl((\T_{20}^{[i+1]})^\top \Q^{-1} (\T_{10}^{[i+1]})(\T_{11}^{[i+1]})(\T_{12}^{[i+1]})\F^{[i]}(\T_{22}^{[i]})^\top \bigr)\bigr) \notag\\ 
& \left. \qquad\times (\T_{22}^{[i]}\I^{[i]}(\T_{22}^{[i]})^\top)^{\dagger}\right)\notag\\
\T_{22}^{[i+1]}&=\text{ReLu}\left(\bigl((\T_{21}^{[i+1]})^\top(\T_{20}^{[i+1]})^\top\Q^{-1}(\T_{20}^{[i+1]})(\T_{21}^{[i+1]})\bigr)
 \times \bigl(\bigl((\T_{21}^{[i+1]})^\top(\T_{20}^{[i+1]})^\top\Q^{-1}\A^{[i]}\bigr)\notag \right.\\ 
& \left. \qquad  -\bigl((\T_{21}^{[i+1]})^\top(\T_{20}^{[i+1]})^\top\Q^{-1} 
\times (\T_{10}^{[i+1]})^\top(\T_{11}^{[i+1]})^\top(\T_{12}^{[i+1]})^\top\F^{[i]}\bigr)\bigr)(\I^{[i]})^{-1}\right)\notag\\
\D_0^{[i+1]} & =\text{ReLu}\left( \B^{[i]} (\D_2^{[i]})^\top (\D_1^{[i]})^\top (\D_1^{[i]} \D_2^{[i]} \Sigmab^{[i]} (\D_2^{[i]})^\top (\D_1^{[i]})^\top)^{\dagger} \right)\notag\\
\D_1^{[i+1]} & =  \text{ReLu}\left(((\D_0^{[i+1]})^\top \R^{-1} \D_0^{[i+1]})^{\dagger} ((\D_0^{[i+1]})^\top \R^{-1} \B^{[i]} (\D_2^{[i]})^\top \right.\nonumber\\
      &\left. \qquad \times (\D_2^{[i]} \Sigmab^{[i]} (\D_2^{[i]})^\top)^{\dagger}\right) \notag\\
    \D_2^{[i+1]} & = \text{ReLu} \left(((\D_1^{[i+1]})^\top (\D_0^{[i+1]})^\top \R^{-1} \D_0^{[i+1]} \D_1^{[i+1]})^{\dagger} (\D_1^{[i+1]})^\top \right.\nonumber\\
    &\left. \qquad \times (\D_0^{[i+1]})^\top \R^{-1} \B^{[i]} (\Sigmab^{[i]]})^{-1}\right). 
    \label{eq:Mstepupdate}
\end{align}}}
\normalsize
\cblue{Hereabove, we use pseudo-inverse operator denoted by $(\cdot)^\dagger$. Each operator is passed over activation function ReLu which stands for, $\text{ReLu}\left(\cdot\right)$ the rectified linear unit function, that projects each entry of its input to the positive orthant. } 

\subsection{\cmazenta{Model Summary}}

The Proposed algorithm is summarized in
Alg.\ref{algo:DeCrypt}. The algorithm infers the probabilistic estimation of the hidden state $(\z_k)_{1 \leq k \leq K}$ jointly with the estimation of latent spaces $\{\T_{10},\T_{11},\T_{12},\T_{20},\T_{21},\T_{22},\D_0,\D_1,\D_2\}$ by following the eq. \eqref{eq:model}. The DeCrypt model is ran for $i_{\max}$ number of iteration to achieve the stabilisation of the latent spaces.

\cblue{
\begin{table}[h!]
\vspace{4mm}
    \centering
    \begin{tabular}{|p{1.1\columnwidth}|}
    \hline
\begin{Algoritmo}
\label{algo:DeCrypt}
\cblue{
\textbf{\cmazenta{Proposed model inference algorithm.}}
\begin{enumerate}
  \item[] \textbf{Inputs.} Prior parameters $(\overline{\z}_0,\P_0)$ ; model noise covariance matrices $\Q$, $\R$ ; set of observations $\{\x_k \}_{1 \leq k \leq K}$ and control input $(\u_k)_{1 \leq k \leq K}$  .
\item[] \textbf{Initialization.} Set positive latent factors
\item[] $\{\T_{10},\T_{11},\T_{12},\T_{20},\T_{21},\T_{22},\D_0,\D_1,\D_2\}$.
\item[] \textbf{Recursive step.} For $i=0,1,\ldots,i_{\max}$: 
 \begin{enumerate}
	 \item[(E step)] Run the Kalman filter  \eqref{eq:predict}-\eqref{update_kalman} and RTS smoother   \eqref{eq:smooth} using latent factors~$\T_{10}^{[i]}$, $\T_{11}^{[i]}$, $\T_{12}^{[i]}$, $\T_{20}^{[i]}$, $\T_{21}^{[i]}$, $\T_{22}^{[i]}$, $\D_0^{[i]}$, $\D_1^{[i]}$, $\D_2^{[i]}$.
	\item[] Calculate ${\Sigmab^{[i]},\Phib^{[i]},\B^{[i]},\C^{[i]},\Deltab^{[i]},\A^{[i]},\F^{[i]},\I^{[i]}}$ using \eqref{eq:RTSquantities}.
	\item[(M step)] Compute
	$\{\T_{10}^{[i+1]},\T_{11}^{[i+1]},\T_{12}^{[i+1]}, \T_{20}^{[i+1]},\T_{21}^{[i+1]},\T_{22}^{[i+1]},  \D_0^{[i+1]},\D_1^{[i+1]},\D_2^{[i+1]}\}$
	 using \eqref{eq:Mstepupdate}.
 \end{enumerate}
\item[] \textbf{{Output.}} State filtering/smoothing pdfs \eqref{Kal_output} and \eqref{eq:RTSoutput} along with pointwise estimates of the latent factor from \eqref{eq:Mstepupdate}.
\end{enumerate}
}
\end{Algoritmo}\\
        \hline
\end{tabular}
\end{table}
}
\cblue{
\subsubsection{Time complexity}
$\rhd$ \textbf{Training.}
We describe the time complexity for training DeCrypt in a given window of length $\tau$. The complexity can be understood by delving into Kalman-based approaches \citep{montella2011kalman}, which imply complexity analysis to $\mathcal{O}(\tau N_z^{2.376})$.\\ 
$\rhd$ \textbf{Testing.}
In testing phase we just perform evaluation of multi-linear equation (Equation \ref{eq:xhat}). This concludes the complexity of $\mathcal{O}(N_x N_z^2)$ for each window. This can be further optimized to $\mathcal{O}(N_z^2)$  if performing forecast for just one feature (which looks very similar to the proposed case, ie. forecasting close price.)}
\normalsize
\section{Application to cryptocurrency price forecasting}
\cmazenta{This section discuss in detail how the proposed approach is applied on the very challenging application of predicting next day prices for crypto-currency.}
\subsection{Training}
\label{sec:training}

The major drawback of the \cmazenta{Expected-Maximization (EM)} strategy used in the proposed approach is that it requires reprocessing on {\em the entire sequence} to estimate the state, control input, and observation transition operators / matrices. The approach requires imposing explicit prior static assumptions on these parameters for the entire duration of the sequence. Such an assumption may not be an appropriate in practice owing to the volatility of the data; furthermore, processing the entire sequence will be computationally expensive. Owing to the volatility, the parameters should be given the freedom to learn and evolve with time. We thus propose an online implementation based on a simple windowing strategy. Thus we are able to relax the non-volatility assumption on the entire sequence and reduce processing times.   

\cblue{In the said strategy, a window of size $\tau$ is slid on the entire dataset. For every time stamp $k$, the matrices in the multi-linear operators are estimated using the last $\tau$ observations contained in the set  $\mathcal{X}_k = \{\x_j\}_{j=k-\tau+1}^{k}$ and $\mathcal{U}_k = \{\u_j\}_{j=k-\tau+1}^{k}$. The proposed EM algorithm is iteratively applied on the window to update the state and operators till convergence. Such a strategy reduces the operational cost significantly. Note that the non-volatility assumption is still there, but only on a small window - this is a reasonable assumption. The major challenge is to estimate a reasonable size of $\tau$. A smaller size would be a better approximation for the non-volatility assumption but would lead to over-fitting on the multi-linear model. On the other hand a larger size would be less prone to over-fitting but would be computationally costly. We initialize the matrices of the multi-linear model using the warm start strategy. The matrices for the current window are initialized with the final values of the prior window. Similarly, the mean and covariance are initialized for $k-\tau+1$ with the past information on the operators from the smoothing process.}
\subsection{Forecasting}

As described in detail in Section \emph{Model Details}, we follow the sliding window strategy. Training each observed window $\x_k$ along with control-input $\u_k$, extracts latent space features and helps in updating the parameters by following EM alternately. 
Once EM iteration stabilizes ( \cblue{we have used 50 iterations for EM to converge used in Alg.\ref{algo:DeCrypt}}), we use the latent space features and learned matrices to estimate the close price for the next timestamp (i.e., the day indexed as $k + \tau + 1$).
 \begin{equation}
  \left\{ 
  \begin{array}{ll}
      \z_{k} &= \T_{10}\T_{11}\T_{12} \z_{k-1} + \T_{20}\T_{21}\T_{22} \u_{k} + \v_{1,k},   \\
      \x_k &= \D_0\D_1\D_2 \z_k + \v_{2,k},
    \end{array}
    \right.
    \nonumber
    \end{equation}
where input $\u_{k}$ is $(\u_j)_{k \leq j \leq k + \tau} \in \mathbb{R}^{1}$ which is computed using the technical indicator SMA(simple moving average)\footnote{\url{https://www.investopedia.com/terms/s/sma.asp}}.\cblue{ Simple moving average (SMA) calculates the average of the a fixed range of prices, usually closing price by the number of period in that range. 
\begin{equation}
    SMA= \frac{c_1 + c_2 + ... + c_n}{n} 
\end{equation}
where $c_n$= closing price of an asset for period n and n= number of total periods}\\
The processing sequence(observed) $\x_{k}$ is $(\x_j)_{k \leq j \leq k + \tau} \in \mathbb{R}^{5}$ for every $k \in \{0,\ldots,K-\tau\}$, where $x_j[1]$ is the daily opening price, $x_j[2]$ is the daily adjusted close price, $x_j[3]$ is the daily high value, $x_j[4]$ is the daily low value, and $x_j[5]$ is the daily net asset volume. Running the DeCrypt model on the considered window yields the mean estimate of the five features for the immediate time stamp which can be indexed as $k + \tau + 1$:
\begin{equation}
\widehat{\x}_{k+\alpha +1} = \D_0 \D_1 \D_2 \z_{k+\alpha}^{-},
\label{eq:xhat}
\end{equation}
The associated covariance matrix is defined as $\S_{k+\tau}$ for immediate next time stamp indexed as $k + \tau+ 1$. /cblue{In particular, \emph{DeCrypt} is designed to forecast the whole five dimensional vector, but we focus primarily on prediction of single entry of the vector i.e., adjusted closing price of the sequence.}

\subsection{Uncertainty Quantification}
\label{sec:prob}
The proposed approach is based on a probabilistic framework and hence can provide confidence score/uncertainty quantification associated with each point-wise estimation. The quantification provides predictive distribution of the future observation which is conditioned on previously seen control-input $(\u_k)$ and observed sequence $(\x_k)$. The probabilistic validation provides informed decision about the (un)certainty associated with model estimation while predicting the future prices of the cryptocurrencies.  
For each index $k$, the distribution of the prediction conditioned on $\widehat{\x}_{k}$ past observations and control-input $\widehat{\u}_{k}$ : 
\begin{equation}
p(\widehat{\x}_{k}|\x_{1:k-1},\u_{1:k-1}) = \mathcal{N}\left(\widehat{\x}_{k};\D_0 \D_1 \D_2 \z_{k}^-,\S_{k}\right),\label{eq_pred}
\end{equation}
where the covariance is defined as $\S_k = \D_0 \D_1\D_2 ((\T_{10} \T_{11} \T_{12}) \P_{k-1} (\T_{10} \T_{11} \T_{12})^\top + \Q) +\R$. It is important to note that $\z_{k}^-$ and $\P_k$ are achieved from Kalman filter, defined in Section E-step in DeCrypt model. The main objective of the proposed approach is to estimate the uncertainty score associated with the prediction given by model for price forecasting. In particular, the main aim is to focus on forecasting the sum of prediction i.e., estimating the first entry $\widehat{\x}_{k}$ denoted by $\widehat{\x}_k[0]$. The quantification about the prediction can be obtained from first row and column of $\S_k$, depicted by $\S_{k}[0,0]$.
We define the (un)certainty score about an increase of the price forecasting value as : 
\begin{align}\label{eq:prob}
    \widehat p_{k} & = \int_{\widehat{\x}_{k}[0]}^{+\infty} \mathcal{N}\left(y;\left[\D_0 \D_1 \D_2 \z_{k}^-\right][0],\S_{k}[0,0]\right)dy  \\
    & = 1 - \text{CDF}(\widehat{\x}_{k}[0]| \left[\D_0 \D_1 \D_2 \z_{k}^-\right][0],\S_{k}[0,0]),
\end{align}
where CDF depicts the cumulative distribution function for the multivariate predictive distribution. The equation above quantifies and estimates the probability that forecasting price value will grow in the future time stamp. After estimating $\widehat{p}_{k}$ for every index $k$, we evaluate cross entropy loss defined as :
\begin{equation}\label{eq:logloss}
    \text{log-loss} = \frac{1}{K}\sum_{k=1}^K -\left( L_k [i] \log( \widehat{p}_{k})\right),
\end{equation}
where ground-truth is depicted as $L_k \in \{0,1\}$ at time $k$ to increase/decrease. 
\section{Experiments and Results}
\cmazenta{This section discuss the experimental results with proposed approach and other state-of-the art methods. The section presents qualitative and quantitative analysis, hence presenting comprehensive study for modeling time series signals.}
\subsection{Dataset description} \label{sec:dataset}
In this study, we consider a cryptocurrency dataset comprising of ten cryptocurrencies. The dataset is extracted from the Yahoo finance cryptocurrency repository using Yahoo finance API \footnote{\url{https://finance.yahoo.com/cryptocurrencies}}. It consists of active cryptocurrencies ranging from some old and new cryptocurrencies. The data extracted is about eight years (between 01/01/2014 to 01/06/2021) for Litecoin, Namecoin, Dogecoin, Peercoin, Bitcoin, Ripple, NXT. For Gridcoin and Ethereum, we extracted seven years of data (between 01/01/2015 to 01/06/2021), the time of its first release year. 
The dataset created is divided into train and test datasets. Training data is from the year 2014 to 2018(2017 December). In contrast, testing data is from 2018 to 2021 is used.
\subsection{Baseline methods}
\begin{itemize}
\item N-Beats: Nbeats is a deep neural network structure with forward and backward residual links. It consists of a deep stack of fully connected layers. It does not perform any specific feature engineering or scaling \citep{oreshkin2019n}.
\item \cmazenta{Deep Auto regressive (DeepAR)} : DeepAR functions by producing probabilistic forecast on long time series sequences \citep{salinas2020deepar}. 
\item \cmazenta{Temporal Fusion Transformers (TFT)}: An attention mechanism based recurrent architecture which brings together multi-horizon  forecasting without having prior information on how they interact with the target \citep{tft}. 
\item \cmazenta{Long short term memory(LSTM)} : Stacked LSTM with 2-layer architecture is used to forecast the time series sequences. The LSTM used comprises 50 cells and ReLU activation to estimate the predictions.\citep{elsworth2020time}.
\item \cmazenta{Convolutional neural network- Technical analysis(CNN-TA)} :1-D Time series is converted to 2-D matrix with technical indicators as rows and time-units as columns.  2D CNN is used for classification and regression.\citep{sezer2018algorithmic}
\item \cmazenta{Recurrent dictionary learning (RDL)}: The RDL approach can be assumed to be a shallow (single layer) version of the proposed work. It can only model piece-wise linear functions.\citep{sharma2021recurrent}
\item \cmazenta{Multi filter neural network (MFNN)} : An end-to-end deep neural network comprising of recurrent neural network and convolutional neural network.\citep{long2019deep}
\cblue{\item RAO-ANN: The Rao algorithms can be categorised as the metaphor-less optimization techniques which is utilized in optimizing the parameters of ANN in forecasting crypto-currency prices \citep{nayak2021modeling}}. 
\cblue{\item ARIMA: Autoregressive integrated
moving average (ARIMA) methodology is used to forecast cryptocurrency prices. The paper identifies the parameter of ARIMA model using partial auto-correlation functions (PACF) and auto correlation function (ACF).\citep{abu2017autoregressive}}
\end{itemize}

\cblue{We have compared our proposed approach with above mentioned models. The ARIMA parameters are set to $(p,d,q) = (2,1,2)$ as it was observed to lead to the best practical performance. We modified LSTM from its original version, by removing the softmax layer and instead included a fully-connected layer to obtain a one node output. The Adam optimizer has been used with learning rate of  $10^{-4}$, $200$ epochs and batch-size $16$ is maintained to minimize the root mean square error. For methods like N-Beats, DeepAR, TFT, CNN-TA, MFNN, RDL, RAO-ANN we stick to their original implementation as described in respective papers mentioned above.}
\begin{figure}[!ht]
\centering
\begin{tabular}{@{}c@{}c@{}}
{\includegraphics[width =0.5\linewidth,height=5cm,keepaspectratio]{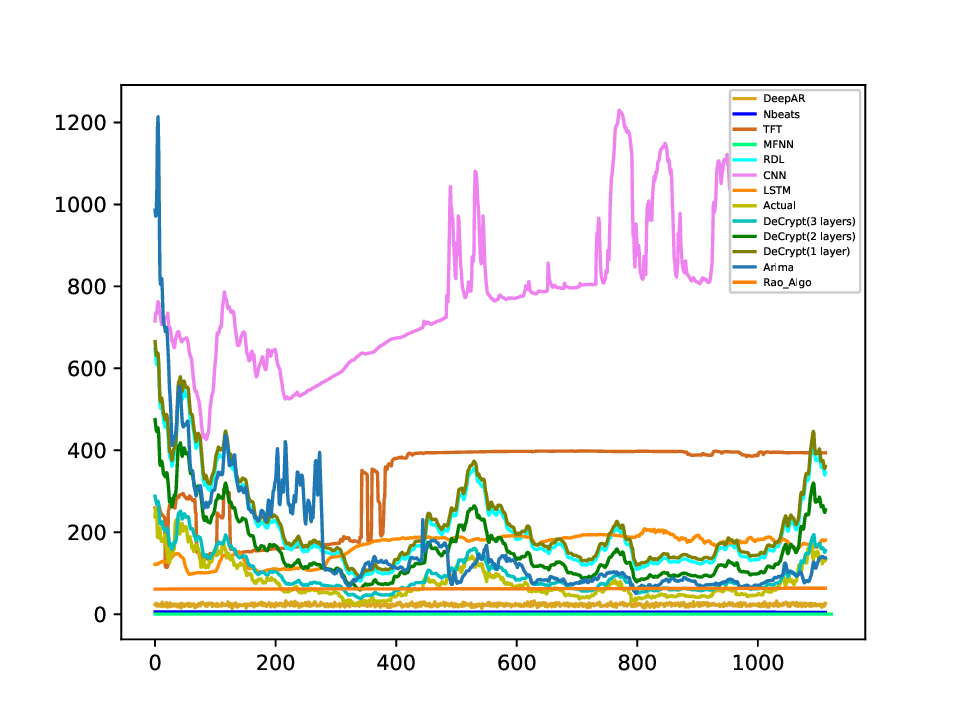}}
&{\includegraphics[width=0.5\linewidth,height=5cm,keepaspectratio]{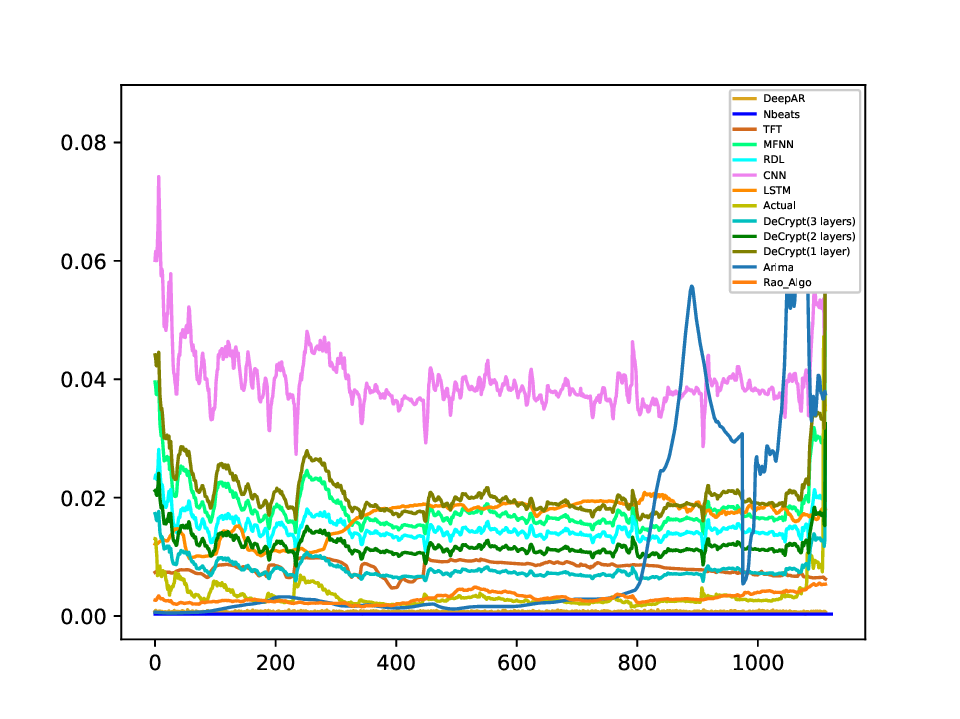}}\\
(a) \texttt{Litecoin} & (b) \texttt{Dogecoin} \\
{\includegraphics[width=0.5\linewidth,height=5cm,keepaspectratio]{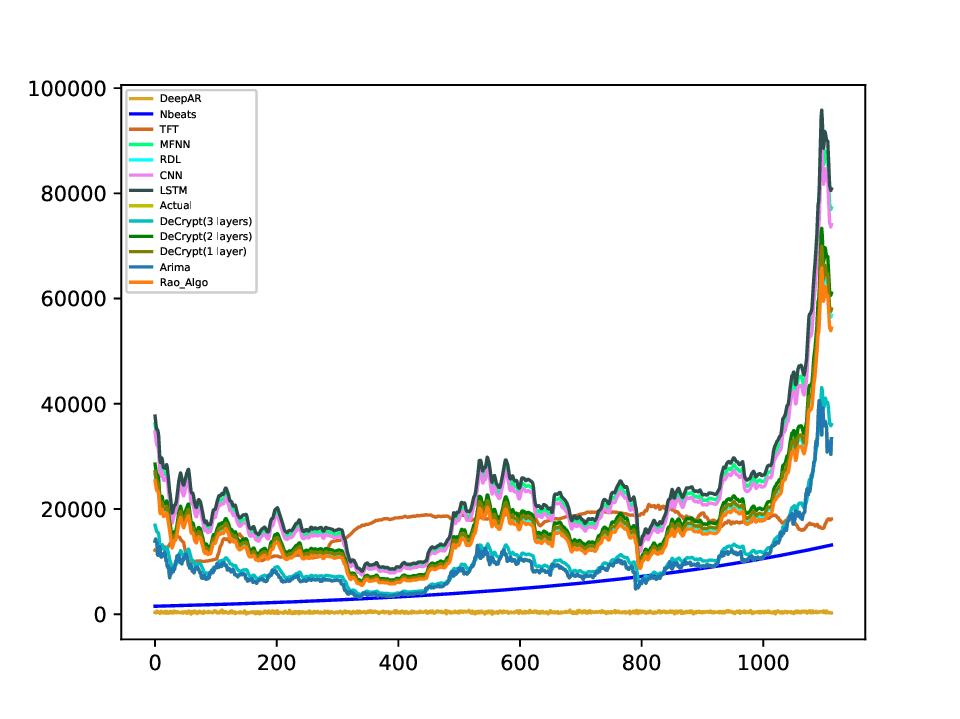}} &{\includegraphics[width=0.5\linewidth,height=5cm,keepaspectratio]{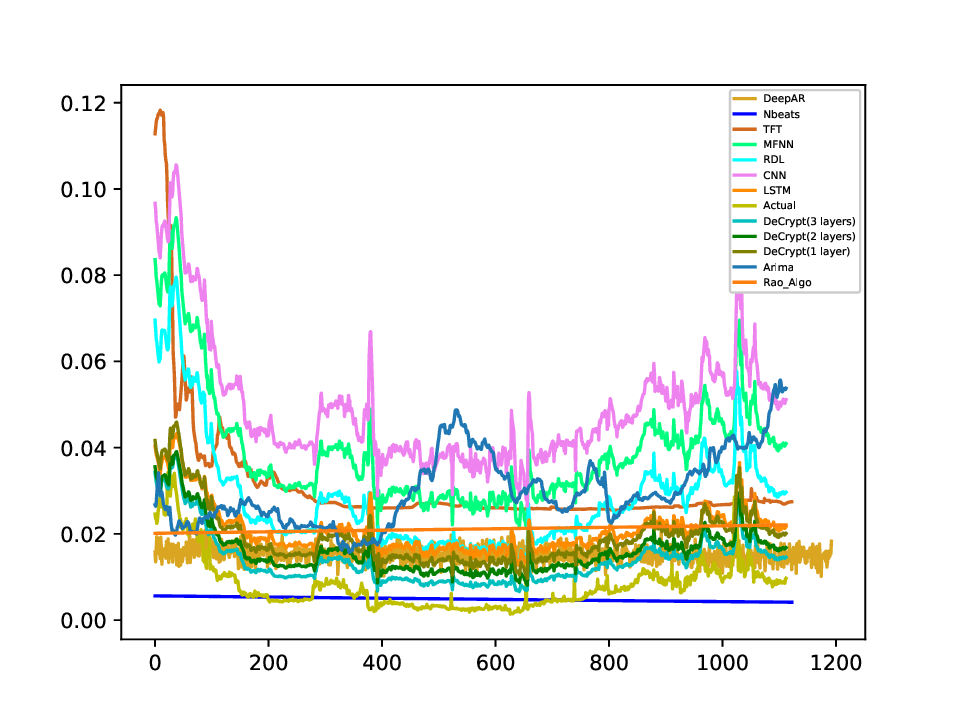}}\\
(c) \texttt{Bitcoin} & (d) \texttt{Gridcoin}\\
\end{tabular}
\small{
\caption{ Cryptocurrency price forecasting via different algorithms \cblue{evaluating test data for}  (a) Litecoin, (b) Dogecoin, (c) Bitcoin, (d) Gridcoin}}
\label{fig:prediction}
\end{figure}
\begin{figure}[!ht]
\centering
\begin{tabular}{@{}c@{}c@{}c@{}}
{\includegraphics[width =0.33\linewidth,height=10cm,keepaspectratio]{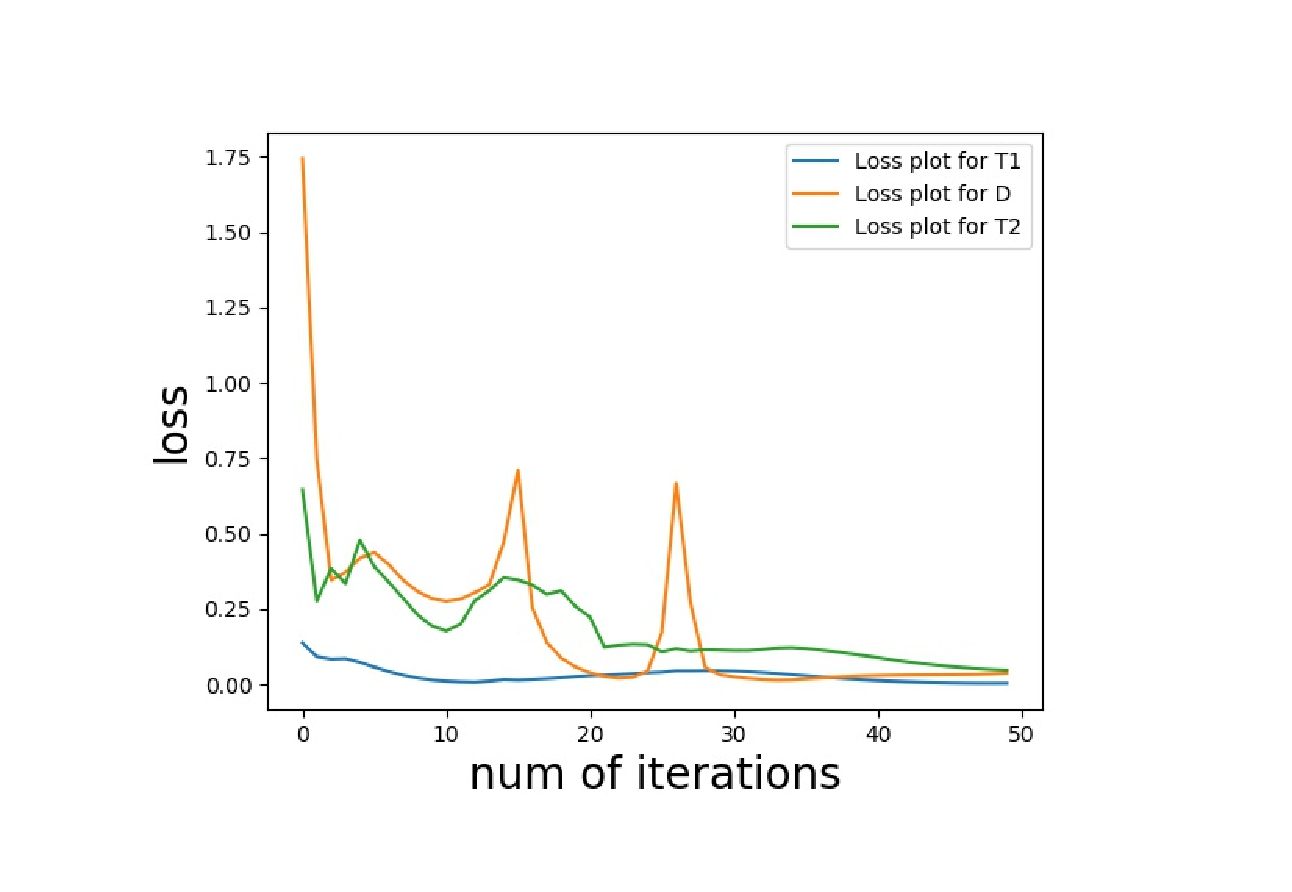}}
&{\includegraphics[width=0.33\linewidth,height=10cm,keepaspectratio]{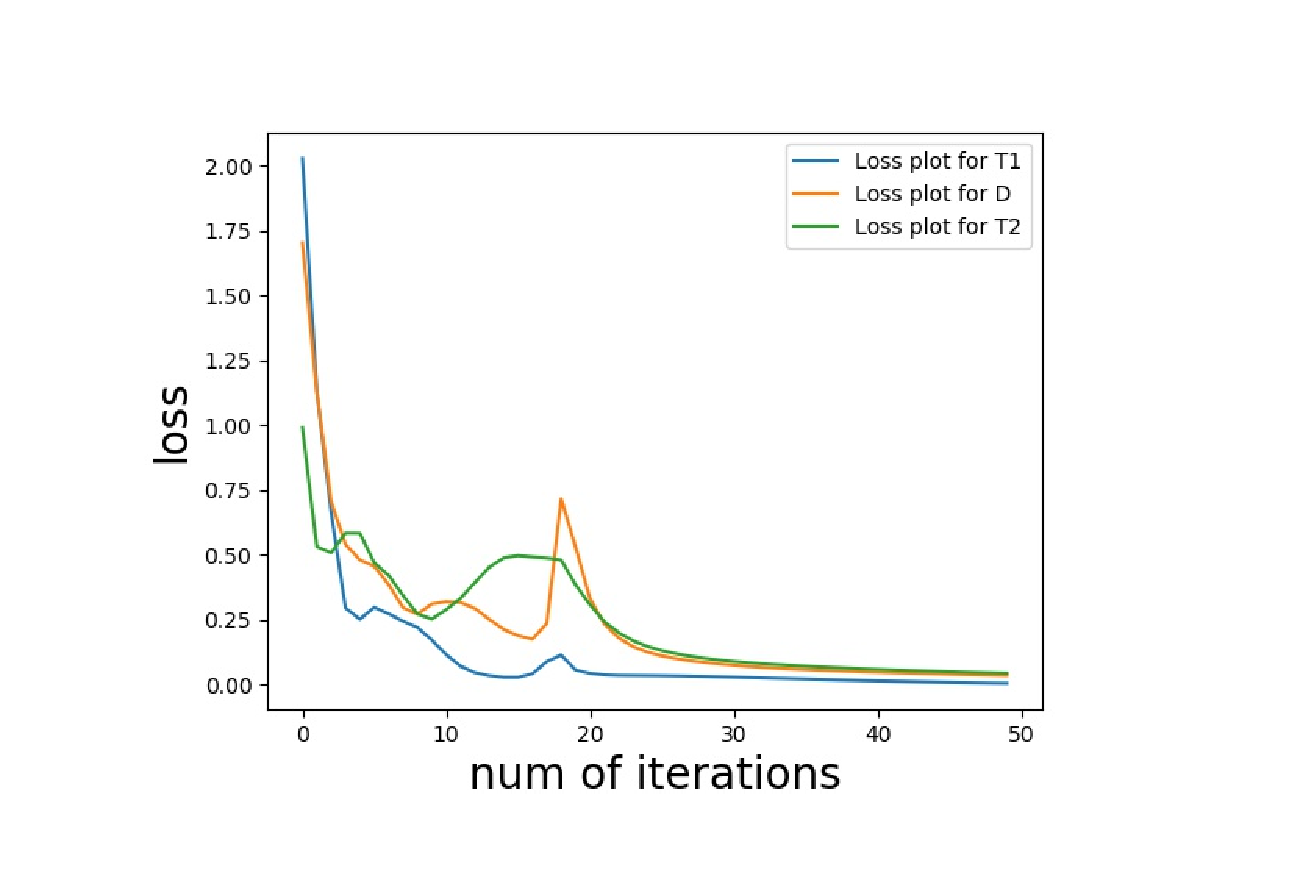}}
&{\includegraphics[width=0.33\linewidth,height=10cm,keepaspectratio]{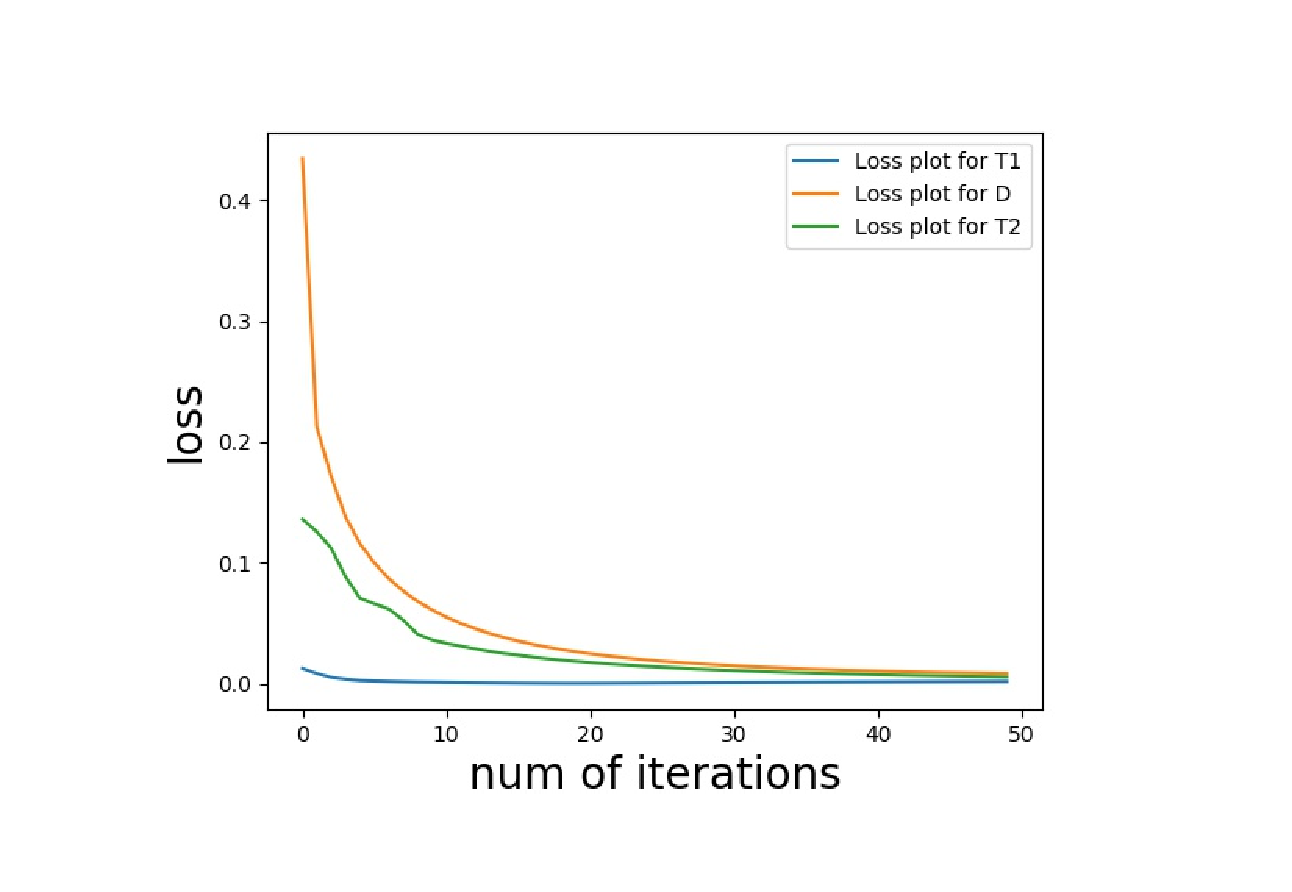}}\\
(a) \texttt{1-Layer} & (b) \texttt{2 Layers} & (c) \texttt{3 Layers} \\
& (A.) Convergence plot for Bitcoin &  \\
{\includegraphics[width =0.33\linewidth,height=10cm,keepaspectratio]{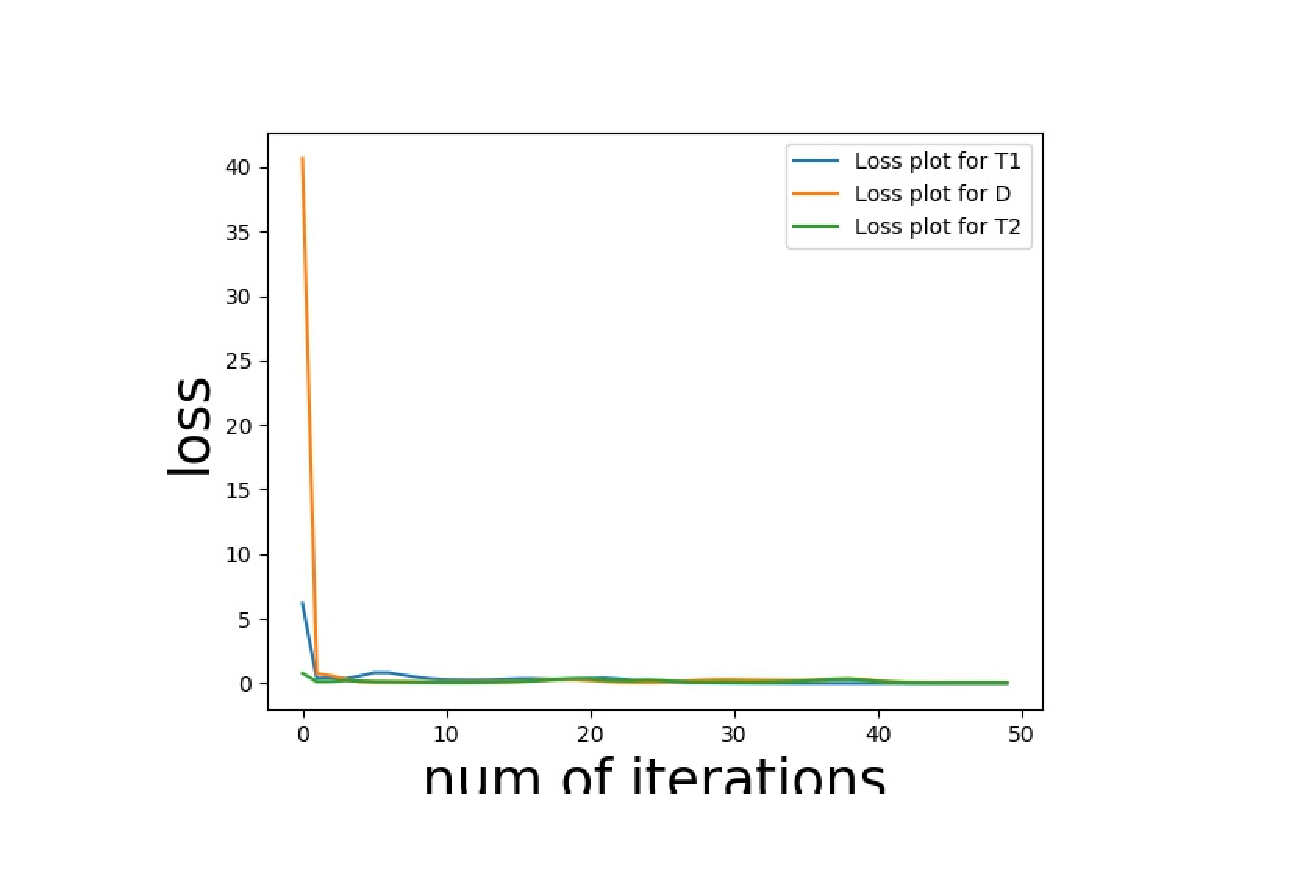}}
&{\includegraphics[width=0.33\linewidth,height=10cm,keepaspectratio]{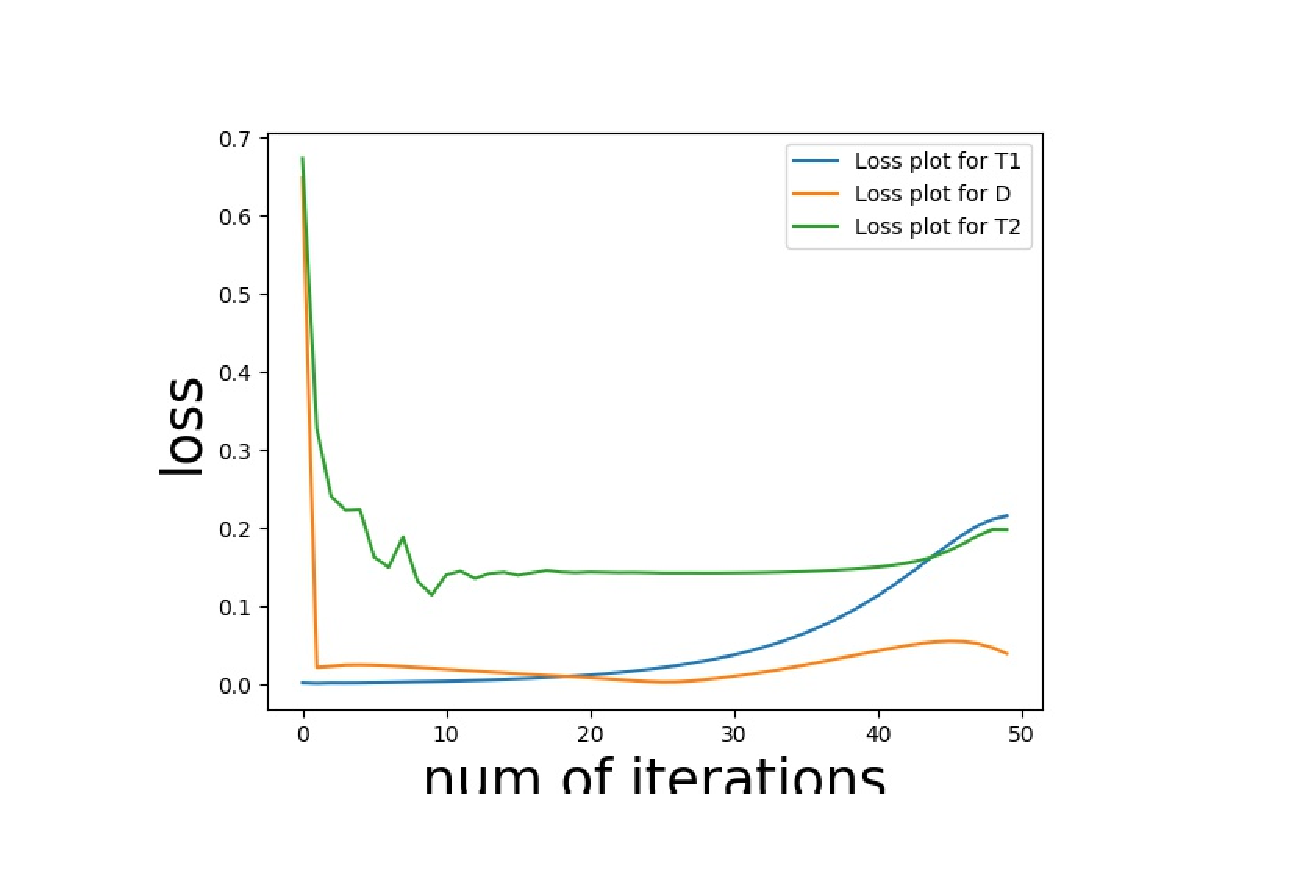}}
&{\includegraphics[width=0.33\linewidth,height=10cm,keepaspectratio]{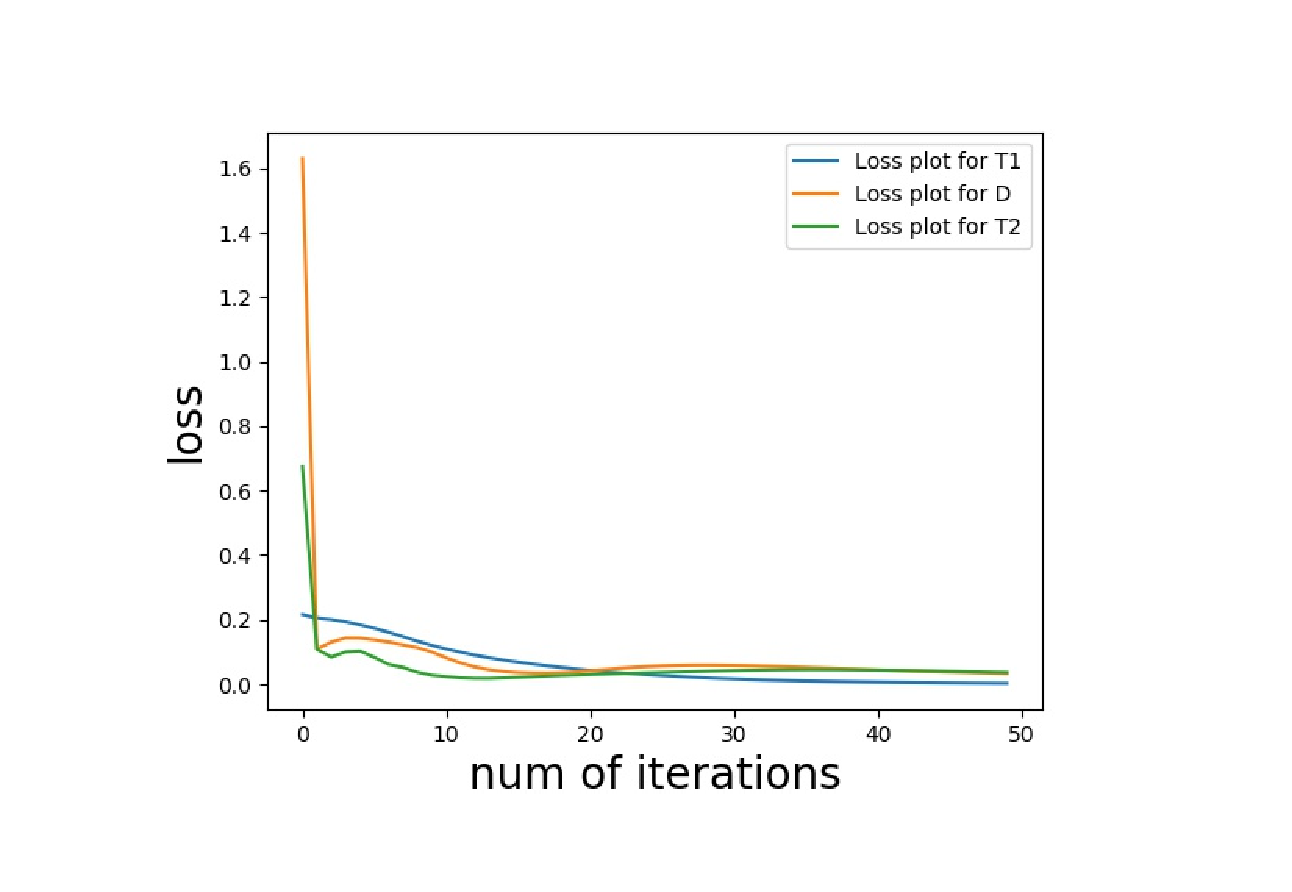}}\\
(a) \texttt{1 Layer} & (b) \texttt{2 Layers} & (c) \texttt{3 Layers} \\
& (B.) Convergence plot for Gridcoin &  \\
{\includegraphics[width =0.33\linewidth,height=10cm,keepaspectratio]{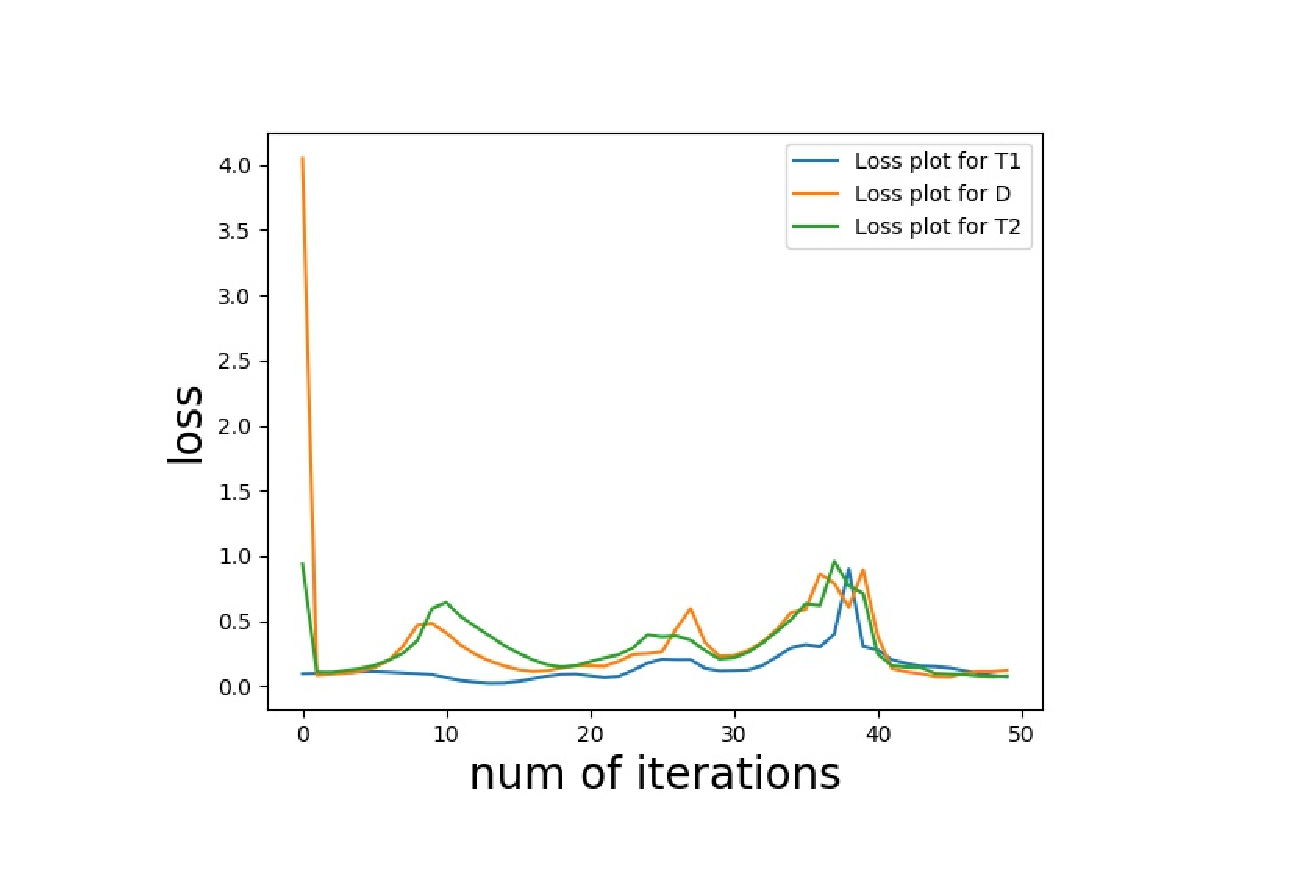}}
&{\includegraphics[width=0.33\linewidth,height=10cm,keepaspectratio]{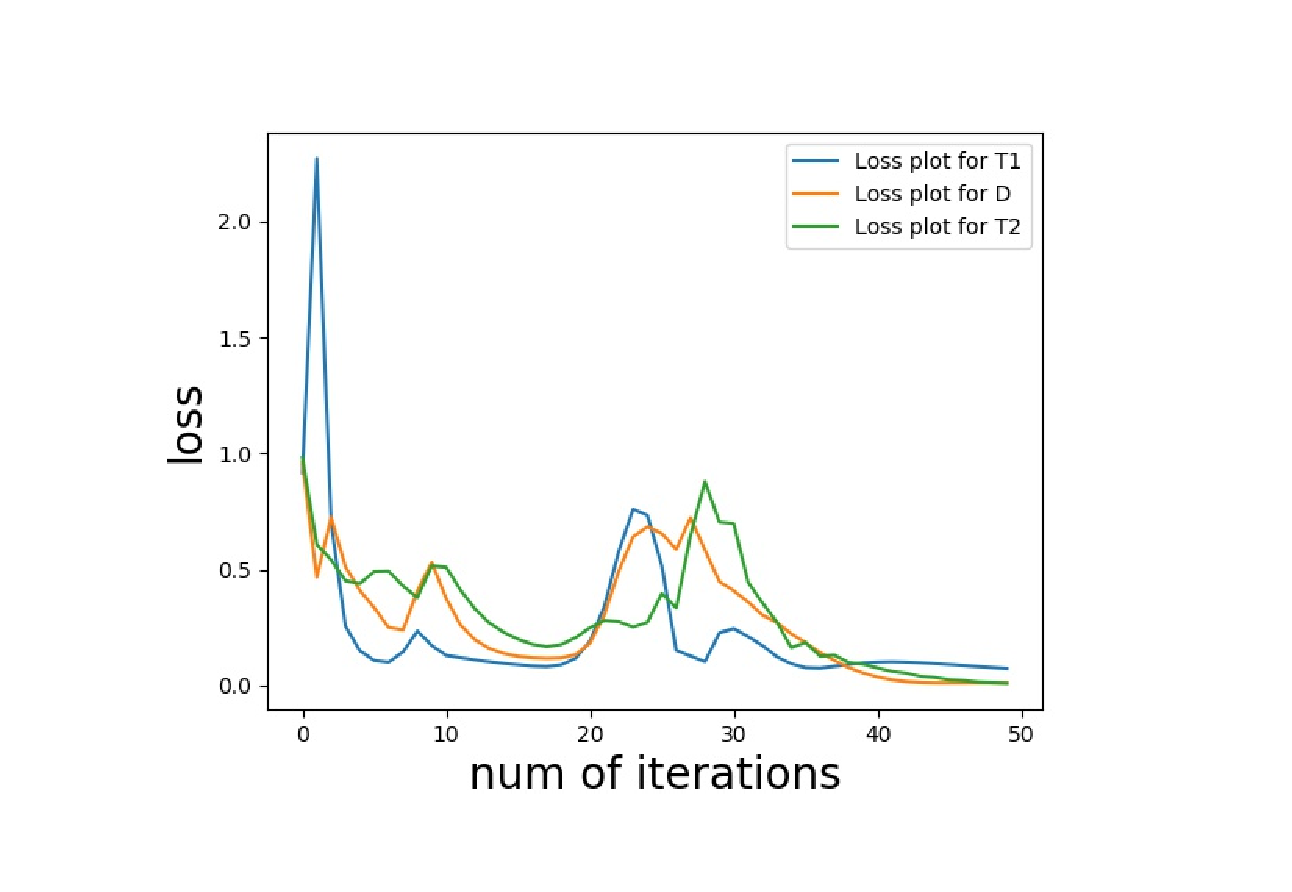}}
&{\includegraphics[width=0.33\linewidth,height=10cm,keepaspectratio]{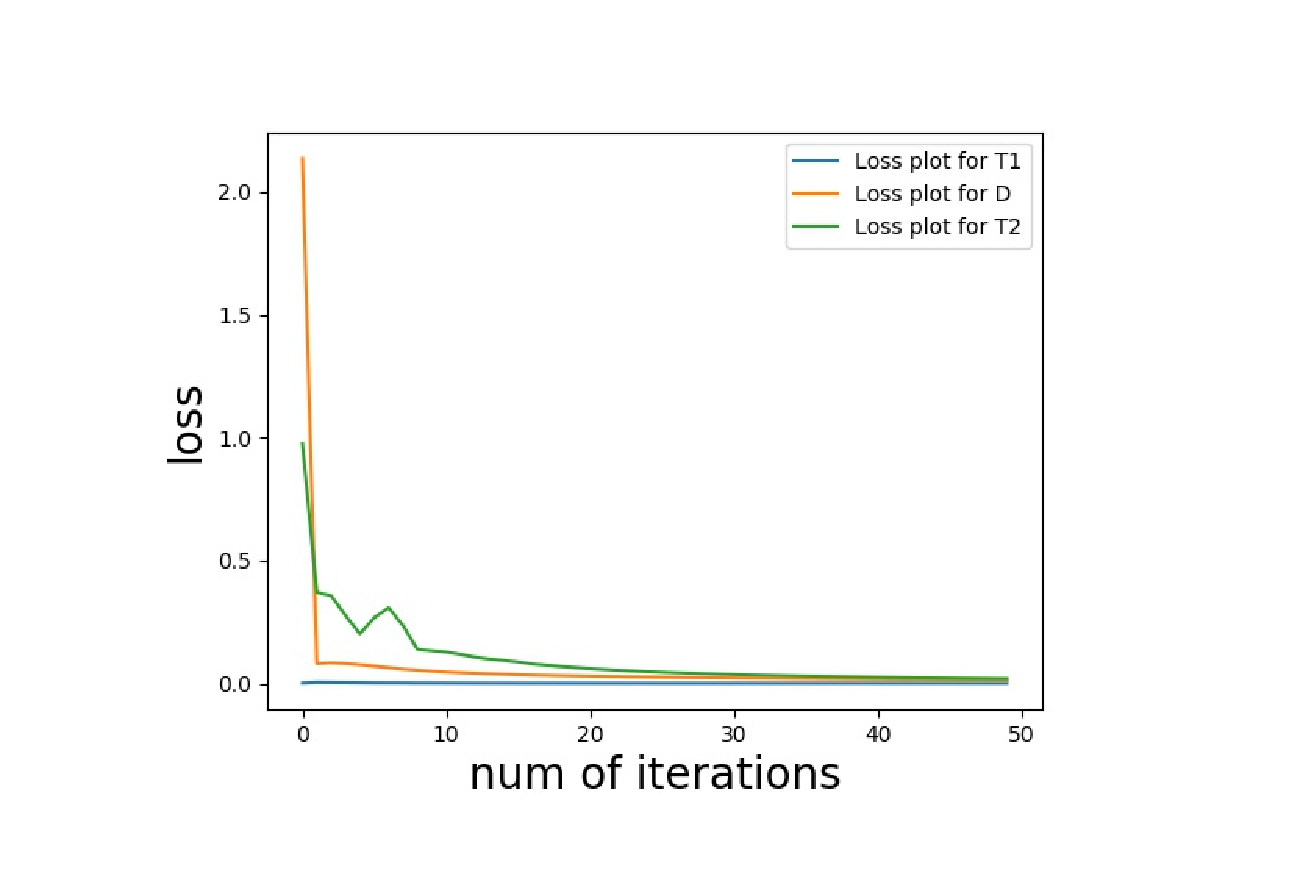}}\\
(a) \texttt{1 Layer} & (b) \texttt{2 Layers} & (c) \texttt{3 Layers} \\
& (C.) Convergence plot for Litecoin &
\end{tabular} 
\caption{ Convergence plots for different layer architecture((a.) 1 Layer, (b.) 2 Layers, (c.) 3 Layer) for (A.) Bitcoin, (B.) Gridcoin, (C.) Litecoin}
\label{fig:Convergence_plots}
\end{figure}
\begin{figure*}[!ht]
\centering
{\includegraphics[width =\linewidth,keepaspectratio, trim = 0cm 7cm 0cm 1cm, clip]{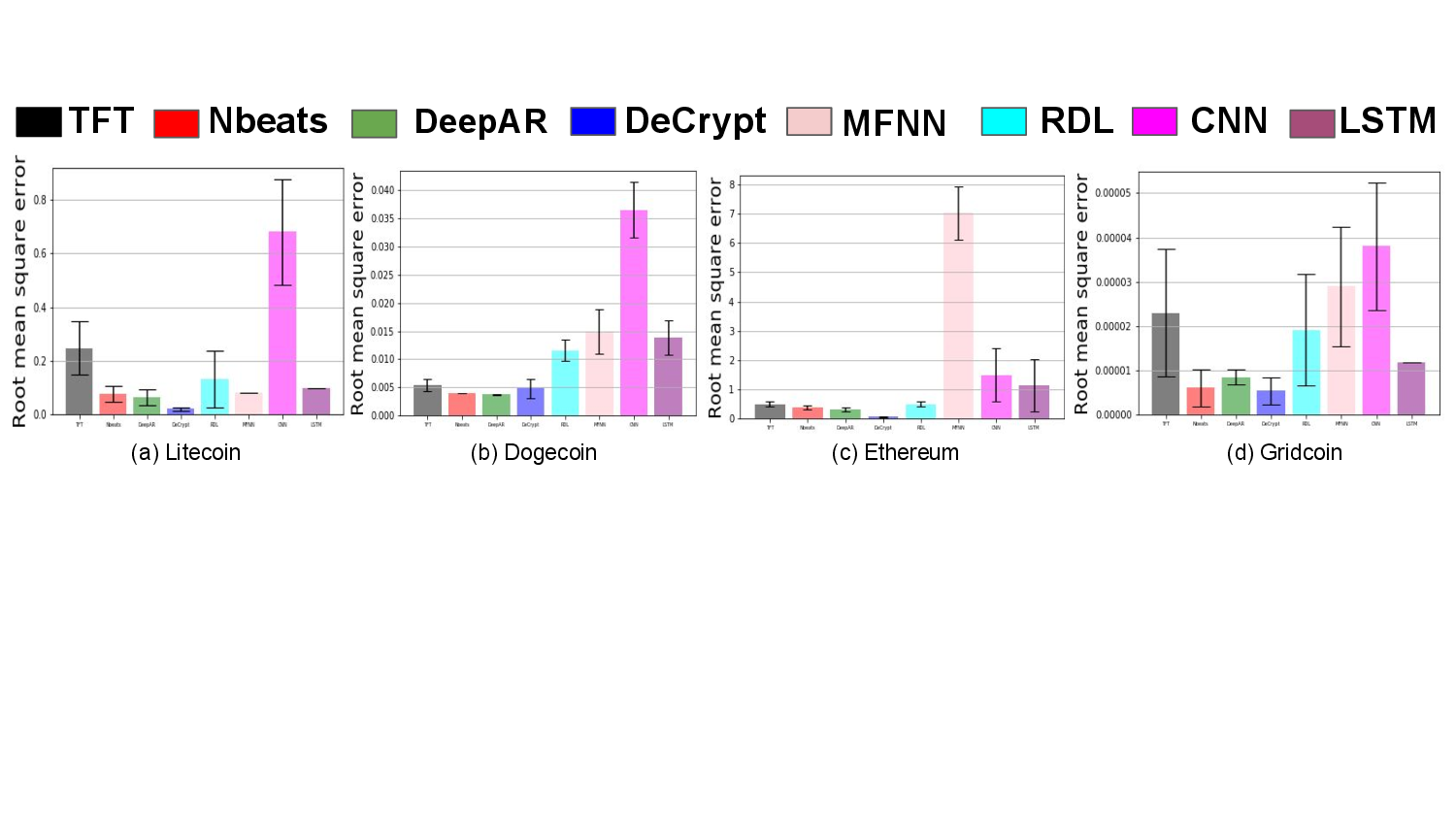}
\caption{ Error bar plot comparing RMSE for DeCrypt against other baseline methods for (a) Litecoin, (b) Dogecoin, (c) Ethereum, (d) Gridcoin.}}
\label{fig:error_bar}
\end{figure*}
\subsection{\cmazenta{Proposed model Parameter analysis}}
The proposed approach is non-parametric. It only requires the specification of the window size; we found that $\tau=50$ yields good results for all cryptocurrencies. The rest of the variables require initialization. These are given below:
$\P_0 = \sigma_P^2 \mathbf{I}$ 

$\Q = \sigma_Q^2 \mathbf{I}$ 

$\R = \sigma_R^2 \mathbf{I}$

$(\sigma_Q,\sigma_R,\sigma_P) = (10^{-5},10^{-1},10^{-1})$, $\mathbf{I}:$ identity matrix 

$\overline{z}_0 = \mathbf{0}$

$N_z$= 5,$N_y$= 1, $N_x$=5,

During training, the entries of the \cmazenta{Deep neural network (DNN)} matrices / operators are initialized at time $0$ using a uniform distribution on $[0,10^{-1}]$. During the test phase, these matrices are fixed, and only the Kalman/RTS inference is run. All presented scores are averaged over $10$ trials, and computed only during the test phase.
More specifically, we will distinguish in the experiments:
\begin{description}
\item[DeCrypt (1 layer):] $\T_{11} = \T_{12} =\rm{\bf{I}}$ is fixed and $\{\T_{10}\}$ is estimated; $\T_{21} = \T_{22} =\rm{\bf{I}}$ is fixed and $\{\T_{20}\}$ is estimated;
and $\D_1 = \D_2 =\rm{\bf{I}}$ fixed and $\{\D_0\}$ is estimated;
\item[DeCrypt (2 layers):] $\T_{12} =\rm{\bf{I}}$ is fixed and $\{\T_{10}\,\T_{11}\}$ is estimated; $\T_{22} =\rm{\bf{I}}$ is fixed and $\{\T_{20},\T_{21}\}$ is estimated;
and $ \D_2 =\rm{\bf{I}}$ fixed and $\{\D_0\,\D_1\}$ is estimated;
\item[DeCrypt (3 layers):] $\{\T_{10}\,\T_{11}\,\T_{12}\}$, $\{\T_{20}\,\T_{21}\,\T_{22}\}$ is estimated $\{\D_0,\D_1,\D_2\}$ are estimated.
\end{description}

\cblue{It is important to note that ignoring the positivity constraints on DeCrypt (1 layer) would identify with the previous work \citep{sharma2021sequential}.}
For benchmarking we have used the commons metrics for regression; they are {\bf Root mean square error} (RMSE),  {\bf Mean Absolute Percentage Error} (MAPE) and {\bf Symmetric Mean Absolute Percentage Error} (SMAPE). All the said metrics are based on the error between the actual and predicted prices and a lower value implies better result. We have also computed the {\bf Pearson correlation coefficient} (r) between the predicted and actual prices; for this metric a higher value implies better result.

We are showing these results for a given window size $\tau=50$ and number of layers. If we continue to increase the window size, the results start to deteriorate, this is likely due to over-fitting. Increasing the window size does not help either; larger window size fails to capture the volatility of the data and hence the performance falls.

\subsection{Result Analysis Discussions}
\cblue{In this section we focus on the performance analysis of the proposed approach. The proposed approach (DeCrypt) has been compared with various numerical methods like N-Beats, DeepAR, TFT, CNN-TA, MFNN, RDL, RAO-ANN, ARIMA, RAO-ANN.}

\subsubsection{\cblue{Influence of window size and depth}}
\begin{table}[!ht]
\small
    \centering
    \caption{\cmazenta{Comparative performance of the proposed approach against different window size for 1-layer architecture.} Lower value($\downarrow$) is considered the better. }
    {%
    \begin{tabular}{|c|c|c|c|c|}
    \hline
        Window size($\tau$) & $r$ & RMSE$\downarrow$ & MAPE(\%)$\downarrow$ & SMAPE(\%)$\downarrow$\\
    \hline
        10 & 0.38 &0.51 & 72.1  & 69.3  \\
        \hline
        15 & 0.37 &0.50 & 73.2  & 70.4  \\
        \hline
        20 & 0.42 &0.47 & 69.6  & 65.2 \\
        \hline
        25 & 0.47 & 0.45 & 65.4 & 61.2  \\
        \hline
        30 & 0.53  & 0.43 & 61.3 & 57.3\\
        \hline
        35 & 0.57 & 0.46  & 58.5  & 53.4 \\
        \hline
        40 & 0.63 &0.39  & 51.8 & 52.3 \\
        \hline
        45 & 0.67& 0.32 & 43.6 & 41.5\\
        \hline
        50 & 0.73& 0.21 &  35.7 & 32.3 \\
        \hline
        55 &0.73 &0.22 & 36.2 & 33.2 \\
        \hline
        60 &0.71 & 0.20& 33.2 & 31.7 \\
    \hline
    \end{tabular}
    }
    \label{tab:result-window1}
\end{table}
\begin{table}[!ht]
\small
    \centering
    \caption{\cmazenta{Comparative performance of the proposed approach against different window size for 2-layer architecture.} Lower value($\downarrow$) is considered the better. }
    {%
    \begin{tabular}{|l|c|c|c|c|}
    \hline
        Window size($\tau$) & $r$ & RMSE$\downarrow$ & MAPE(\%)$\downarrow$ & SMAPE(\%)$\downarrow$\\
    \hline
        10 & 0.42 & 0.59 & 74.1 & 71.2 \\ 
        \hline
        15 & 0.45 & 0.54 & 73.5  &  70.7 \\
        \hline
        20 & 0.49 & 0.43 & 71.8  &70.2 \\
        \hline
        25 & 0.53 & 0.41 & 68.5 & 65.6 \\
        \hline
        30 & 0.57 & 0.39 &  65.1  & 64.3 \\
        \hline
        35 & 0.59 & 0.33 & 58.6 & 57.8 \\
        \hline
        40 & 0.65 & 0.27 & 51.5 & 50.8\\
        \hline
        45 & 0.69 & 0.21 & 47.5 & 39.7 \\
        \hline
        50 & 0.76 & 0.17 & 31.2 & 28.7  \\
        \hline
        55 & 0.77 & 0.17 & 32.3 & 29.6  \\
        \hline
        60 & 0.75 & 0.15 & 31.6 & 27.8 \\
    \hline
    \end{tabular}
    }
    \label{tab:result-window2}
\end{table}
\begin{table}[!ht]
\small
    \centering
    \caption{\cmazenta{Comparative performance of the proposed approach against different window size for 3-layer architecture.} Lower value($\downarrow$) is considered the better. }
    \begin{tabular}{|c|c|c|c|c|}
    \hline
        Window size($\tau$) & $r$ & RMSE$\downarrow$ & MAPE(\%)$\downarrow$ & SMAPE(\%)$\downarrow$\\
    \hline
        10 & 0.43  & 0.59  & 0.76  & 0.72\\
        \hline
        15 &  0.47 & 0.56  & 0.71  & 0.71 \\
        \hline
        20 &  0.48 & 0.51 &  0.67 & 0.65\\
        \hline
        25 & 0.53  & 0.47  & 0.61  & 0.62 \\
        \hline
        30 &  0.59 & 0.42  &  0.59 & 0.57 \\
        \hline
        35 & 0.64  & 0.35  &  0.54 & 0.51 \\
        \hline
        40 & 0.69  & 0.27  & 0.43  &  0.42\\
        \hline
        45 & 0.73  & 0.21  &  0.39 & 0.33 \\
        \hline
        50 & 0.81 & 0.12 & 29.1 & 21.2 \\
        \hline
        55 & 0.82 & 0.14 &  29.3 & 20.3 \\
        \hline
        60 & 0.81 & 0.13 & 28.4 & 20.4 \\
    \hline
    \end{tabular}
    \label{tab:result-window3}
\end{table}
\normalsize
\cblue{ The proposed solution is non-parametric. The only design parameters that need to be fixed are the window size and depth. Therefore, it is very important to choose the optimal window size which finds a good balance between the model complexity (depth) and accuracy. To better understand, we present Table ~\ref{tab:result-window1}, Table ~\ref{tab:result-window2}, Table ~\ref{tab:result-window3} which represent the window size analysis in DeCrypt (1 layer), DeCrypt (2 layers), DeCrypt (3 layers) architecture respectively. The above mentioned tables provides comprehensive analysis on the performance of the model on varying window sizes. A comprehensive study has been compiled which offers analysis of various metrics like Pearson correlation (r),  RMSE (Root Mean Square Error), MAE (Mean Absolute Error), and SMAPE (Symmetric mean absolute percentage error) for different window sizes $\tau$. After analysing the results we conclude that model performance keeps on improving as window size increases till a stabilization point $\tau = 50$. Ideally one would expect that the results would improve as the window size increases; especially for deeper versions. This is because larger window size means more data and  hence better generalisation ability. But note that it is a dynamical model. The tacit assumption here is that in the window the size the underlying dynamical function does not change. While this is true for shorter windows, this does not hold for larger ones as the non-stationarity comes into play. This is the reason we find a trade-off between window size and accuracy. We further set this value of window size in upcoming experiments.}

\subsubsection{Comparison with state-of-the-art methods}

\begin{table}[!ht]
    \centering
    \caption{\cmazenta{Comparative performance of the proposed approach against baseline methods. }Lower value($\downarrow$) is considered the better.}
    {
    \begin{tabular}{|l|c|c|c|c|}
    \hline
        Model & $r$ & RMSE$\downarrow$ & MAPE(\%)$\downarrow$ & SMAPE(\%)$\downarrow$\\
    \hline
        LSTM & 0.33 & 0.71 & 83.2 & 64.3 \\  
        CNN-TA & 0.29 & 0.68 & 91.2  &  70.8 \\
        ARIMA & 0.29 & 0.68 & 91.2  &  70.8 \\
        Rao-ANN& 0.29 & 0.68 & 91.2  &  70.8 \\
        MFNN &  0.31& 0.21  &70.28  &68.2 \\
        N-Beats & 0.38 & 0.27&  41.2& 38.72 \\
        DeepAR &  0.30 &0.39  & 46.8   &41.47 \\
        TFT & 0.52 & 0.24 & 48.95 & 40.62 \\
        RDL &  0.69& 0.20 & 48.7 & 35.68 \\
        \cblue{ARIMA} &  0.58& 0.49 & 56.4. & 46.24 \\
        \cblue{Rao-ANN} &  0.42& 0.59 & 68.7 & 64.38 \\
        \hline
        DeCrypt (1 layer) &0.73  & 0.21 & 35.7 & 32.33 \\
        DeCrypt (2 layers) & 0.76 & 0.17 & 31.2  & 28.7\\
        DeCrypt (3 layers) & {\bf 0.81} & {\bf 0.12} & { \bf 29.14}  & {\bf 21.2}  \\
    \hline
    \end{tabular}
    }
    \label{tab:result-overall}
\end{table}
The overall performance of the model vis-a-vis the state-of-the-art is shown in Table\ref{tab:result-overall}. The Table presents the average results for ten cryptocurrencies; owing to limitations in space we are not able to show individual results. One can see that the proposed method outperforms the baseline by a considerable margin. Fig \ref{fig:prediction} shows the forecast performance of DeCrypt with other baseline methods for visual evaluation and Fig \ref{fig:Convergence_plots} shows the convergence plots for model parameters for different layer of architecture for DeCrypt \cblue{where $\T_1$ is the multi-linear state operator achieved from product of three positive-valued linear factors $\T_{10}\T_{11}\T_{12}$, $\T_2$ is the is the multi-linear control operator achieved from product of three positive-valued linear factors $\T_{20}\T_{21}\T_{22}$ and  $\D$ is the multi-linear observation operator achieved from product of three positive-valued linear factors $\D_0\D_1\D_2$}. It can be clearly seen that the \emph{DeCrypt} with its three layers architecture outperforms the other baseline approaches significantly. This is mainly because of the deeper network's capacity to better model non-linearity compared to the shallower ones. The model achieves a 0.17 points drop in RMSE, 19.56\% drop in MAPE, and 14.48\% drop in SMAPE; it gains 0.12 in Pearson's correlation $r$ compared to the best performing benchmarks. We have plotted the error-bars for four different cryptocurrencies (Litecoin, Dogecoin, Ethereum, Gridcoin)in Fig.\ref{fig:error_bar}.  From these plots the reader can verify that not only is the proposed method more accurate (least mean error) but is also the most robust (least deviation). \cblue{In Table \ref{tab:ttest_score} we present the comparison of performance of proposed approach with state-of-the-art method through statistical test (T-test) with confidence interval of 0.95. We can observe that the T-test values for proposed approach DeCrypt (3 layers) is very small as compared to the other methods, hence we can conclude that more similarity exists between the actual closing prices and predicted closing prices when compared for different Crypto-currencies. Due to space constraints we have provided Avg. Score for t-test for all the ten cryptocurrency for each method in Table \ref{tab:ttest_score}. From the results we conclude that average T-test score is very low for Decrypt (3 layers) method when compared with other state-of-the-art method, hence we conclude that Decrypt (3 layers) performance is very close to ground truth. In contrast when we see individual crypto-currency analysis from table we can see that TFT outperforms all the methods in Gridcoin and DeepAR outperforms Dogecoin.}
\begin{table}[!ht]
\small
    \centering
    {
    \cblue{
    \begin{tabular}{|c||c|c|}
    \hline
{Method} &\multicolumn{1}{|c|}{Train Time cost (h.)}& Test Time cost (min.) \\
  \hline
      \hline
  DeCrypt (3 layers) & {2.21}h& 22 min\\
  \hline
  DeCrypt (2 layers)& {2.32}h & 22.4 min\\
  \hline
  DeCrypt (1 layer)& {1.48}h & 18.8 min\\
  \hline
  ARIMA & 2.31h & 36 min\\
  \hline
  LSTM & 5 days & 41 min\\
  \hline
  DeepAR & 2.45h & 20 min\\
  \hline
  TFT & 2.25h & 27 min\\
  \hline
  Nbeats & 3.12h & 25 min\\
  \hline
  CNN-TA & 4.57h & 40 min\\
  \hline
  MFNN & 4.12h & 37 min\\
  \hline
  RDL & 1.69h & 35 min\\
  \hline
  Rao-ANN & 4.35h & 25 min\\
  \hline
\end{tabular}} }
    \caption{Averaged time over 50 random runs for processing the dataset (train(hrs) and test(min)), for \cmazenta{the proposed approach} and its competitors.}
    \label{tab:time_complexity_methods}
\end{table}
\begin{table}[!ht]
\small
    \centering
    {
    \cblue{
    \begin{tabular}{|c||c|c|c|c|c|}
    \hline
{Method} &\multicolumn{1}{|c|}{Bitcoin}& {Gridcoin}& {Dogecoin}& {Litecoin}& Avg. Score \\
  \hline
      \hline
  DeCrypt (3 layers) & 0.31 & 0.73& 0.82 & 0.57 & 0.68 \\
  \hline
  DeCrypt (2 layers)& 0.56  & 0.85& 0.91& 0.69& 0.72 \\
  \hline
  DeCrypt (1 layer)& 0.58  & 0.98 & 0.99 &0.72 & 0.78\\
  \hline
  ARIMA & 0.90 & 0.87& 0.95 & 0.93 & 0.83\\
  \hline
  LSTM &  0.59& 1.75& 0.97 &0.82 & 0.94\\
  \hline
  DeepAR &  0.54& 0.79 & 0.81& 0.55 & 0.71\\
  \hline
  TFT &  0.40 & 0.70 & 1.12 & 0.71 & 0.73\\
  \hline
  Nbeats & 0.56 & 0.68&0.99& 0.61 & 0.84\\
  \hline
  CNN-TA & 0.87  &1.53 &2.11 &1.19 & 1.13\\
  \hline
  MFNN & 0.82 & 1.30 & 1.21& 0.74 & 1.37\\
  \hline
  RDL & 0.58 & 0.95 & 0.94 &0.68 & 0.84\\
  \hline
  Rao-ANN & 0.54 &0.70 & 1.02& 1.13 & 0.94 \\
  \hline
\end{tabular}}}
    \caption{Comparison of T-test score for \cmazenta{the proposed approach } with state-of-the-art method for (a) Bitcoin, (b) Gridcoin, (c) Dogecoin, (d) Litecoin, (e)Avg. Score for all the ten crypto-currencies.}
    \label{tab:ttest_score}
\end{table}

\normalsize
\cblue{To understand the comparison in performance between the proposed method and state-of-the-art methods, we present Table \ref{tab:time_complexity_methods} which depicts the computational time for forecasting the next day closing price of ten cryptocurrencies. We provide a comprehensive analysis by distinguishing the time required to train and test the methods (on their training and testing time frame as described in sec ~\ref{sec:dataset}) using the walk-forward method described in \citep[Section
4.2.1]{sharma2021recurrent}. We conclude that the highest computational time was consumed by LSTM approach. Among the other methods, DeCrypt (1 layer) and RDL method is the fastest while the computation time of DeCrypt (3 layers) is very much comparable to DeepAR and TFT. However, note that the existing algorithms are optimized to take advantage of GPU, the proposed approach does not, it runs only on the CPU. It may be possible to improve the performance in the future through parallelization.}

\subsubsection{Uncertainty quantification}
\begin{table}[!ht]
\caption{(Un)certainty quantification (log-loss) and Cryptocurrency Volatility Index (CVI) evaluated using DeCrypt (3 layers) and Nbeats}
    \centering
    \small
    {%
    \begin{tabular}{|p{2cm}|p{1.5cm}|p{1.5cm}|p{1.5cm}|p{1.3cm}|p{1.3cm}|}
    \hline
    Cryptocurrency & DeCrypt (1 layer) & DeCrypt (2 layer) & DeCrypt (3 layers) & CVI (DeCrypt*)) & CVI (Nbeats)\\
    \hline
        {Bitcoin} & 0.79  & \bf{0.67}  & 0.86  & \bf{0.31}& 0.48\\ 
        \hline
        {Dogecoin} & \bf{4.32} & 4.89 & 4.75  & \bf{2.62} & 2.71\\
        \hline
        {Namecoin} & 4.34 & 3.79 & \bf{3.65}  & 1.61& 1.81\\
        \hline
        {Litecoin} & 1.21 & 1.10  & \bf{0.92}& \bf{0.35}& 0.49\\
        \hline
        {Gridcoin} & 1.81  & 1.56  & \bf{1.45}  & 0.59 & 0.63 \\
        \hline
        {Peercoin} & 1.67 & 1.43 & \bf{1.23}   & 0.50& 0.59 \\
        \hline
        {Ripple}  &  1.24 & 1.11  &\bf{0.97} &  \bf{0.42} & 0.53\\
        \hline
        {NXT}     & 1.32  & 1.10  & \bf{0.91}  & \bf{0.32} & 0.42\\
        \hline
        {Ethereum}& 1.56  & \bf{1.41}  & 1.43  & \bf{0.38} & 0.63\\
        \hline
        {Binance coin} & 1.34 & 1.21  & \bf{0.84} & \bf{0.33}& 0.72\\
    \hline
    \end{tabular}
}
\label{tab:logloss}
\end{table}
\begin{table}[!ht]
\caption{Cryptocurrency Volatility Index (CVI) evaluated using state-of-the-art method predictions}
    \centering
    \small 
    \cblue{%
    \begin{tabular}{|p{2cm}|p{1cm}|p{1cm}|p{1cm}|p{1cm}|p{1cm}|p{1cm}|p{1cm}|p{1cm}|}
    \hline
    Cryptocurrency & LSTM & CNN-TA & ARIMA & Rao-ANN & MFNN& DeepAR& TFT & RDL\\
    \hline
        {Bitcoin}  &  0.57 &  0.53 & 0.46 &0.59 & 0.72 & 0.52 & 0.49& 0.38\\ 
        \hline
        {Dogecoin} & 2.87  & 3.27  & 2.35& 3.43 &3.87& 2.72& 2.83& 2.57\\
        \hline
        {Namecoin}  & 1.93 & 2.12  & 1.88 & 1.83 & 1.96& \bf{1.58} & 1.68 & 1.73\\
        \hline
        {Litecoin} & 1.31 &  0.56 & 0.48 & 0.51 & 0.67 & 0.38 & 0.46& 0.41\\
        \hline
        {Gridcoin} & 0.99  &  0.87 & 0.67  & 0.74 & 0.96& 0.68 & 0.75 &\bf{0.53}\\
        \hline
        {Peercoin} &  0.81 & 0.78 &\bf{0.48} & 0.69 & 0.93 & 0.64 & 0.71 & 0.58\\
        \hline
        {Ripple}  &  0.78 & 0.83 & 0.64& 0.73 & 0.88 & 0.61& 0.68& 0.54\\
        \hline
        {NXT}   0.57  & 0.63 & 0.46 & 0.53  & 0.64 & 0.58 & 0.54& 0.59& 0.47\\
        \hline
        {Ethereum}& 0.98 &  0.78 &  0.43 & 0.58 & 0.49 & 0.52& 0.61&0.45\\
        \hline
        {Binance coin} &0.53 & 0.64  & 0.41 & 0.49  & 0.51& 0.45& 0.54 &0.39 \\
    \hline
    \end{tabular}
}
\label{tab:CVI_baselines}
\end{table}
The advantage of DeCrypt over other baseline approaches is the estimation of (un)certainty quantification associated with each prediction. For cryptocurrencies this measure will be directly proportional to the volatility index. As discussed in section \emph{Uncertainty Quantification}, it is easy to evaluate (un)certainty of prediction of an increase/decrease of price forecast by calculating the log-loss penalization as explained in eq. \ref{eq:logloss}. To validate the proposed method results on (un)certainty quantification the work also evaluated cryptocurrency volatility index for each cryptocurrency. Cryptocurrecy Volatility Index (CVI) \cblue{\citep{kim2021vcrix,woebbeking2021cryptocurrency}}can be defined as a measure of market's expectation of volatility over the near trading terms for a particular asset. Volatility is often described as the "rate and magnitude of changes in prices" and in finance often referred to as risk. Volatility is sometimes associated with the uncertainty of risk related to the amount of changes in security's value. This can be further described as if the security's value can potentially be spread out over a larger range of values, it indicates that the price of the security can change dramatically over a short time period in either direction which is flagged as higher volatility. On the other hand, A lower volatility means that a security's value does not fluctuate dramatically, and tends to be more steady. The mathematical formula to calculate CVI \citep{woebbeking2021cryptocurrency}:
\begin{equation}
    \text{CVI} = \sqrt{365}*\sqrt{\frac{1}{N}\sum_{N=1}^N ((Close price-Price at N)^2))},
\end{equation}
Table \ref{tab:logloss} depicts the calculated log-loss values for each cryptocurrency vis-a-vis their cryptocurrency volatility index (CVI)\footnote{\url{https://github.com/dc-aichara/PriceIndices}}.\cblue{Table \ref{tab:logloss} represents the log loss score for all the layer architecture for DeCrypt and CVI score for DeCrypt (3 layers) and Nbeats. Table~\ref{tab:CVI_baselines} represents the CVI scores from state-of-the-art method}. A smaller value of the loss should be associated with lower volatility and vice versa. It can be clearly seen that log-loss associated with the Bitcoin, Litecoin, Peercoin, Ripple, NXT, Binance coin is less than one, meaning prices associated with these cryptocurrencies are less volatile. In contrast, Dogecoin, which is highly volatile and has a history of spiked values after a tweet by a major influencer, is more difficult to assess and has a log-loss score of $4.75$. Thus one can see how the proposed algorithm can quantify uncertainty and how this measure is proportionate to the oracle volatility. This is by far the most important result in the paper. This result shows how the proposed approach may be used for practical trading where both the point estimate as well as the uncertainty about the estimate is required for making decisions.

Owing to limitations in space, unfortunately not able to show the convergence of the proposed algorithm or the run-times of different techniques. Although \cmazenta{the work} have used 50 iterations and its empirically checked that the proposed algorithm converges in about 20-25 iterations. The convergence is monotonic. In terms of speed the proposed method is about 2-4 times faster than LSTM, CNN-TA and MFNN, and is about an order of magnitude faster than TFT, Nbeats and DeepAR. Of the existing methods, only RDL is comparable to the proposed method in terms of speed.
\subsection{Discussion}
\cmazenta{Consistently beating in modeling Time series signals has been challenging for a long time. The proposed three-layer architecture method performed better than the current state-of-the-art method. The proposed method is based on the deep state-space and feedback strategy. As can be seen from Fig.~\ref{fig:prediction} the proposed model can predict the sudden spikes in the data compared to other state-of-the-art methods. This is mainly because the proposed method is based on SSM, which uses probabilistic predictive distribution to estimate the future state of the price trajectories. The technique also embeds deep non-negative factors to learn the model parameters. Deep factors helps in updating the model parameters and state-space of unseen signals continuously with time, in contrast to machine learning models, which use a huge amount of data to learn approximations. We observed that when the time series signals grow, these models suffer from vanishing gradient and exploding gradient problems, which hampers learning model parameters. Due to this, model approximations are not learned efficiently and cannot capture sudden spikes (highs and lows in prices). These models also suffer from over-fitting. The proposed method avoids over-fitting as we move ahead in the sliding window protocol, and previously updated model parameters are used as initialized values for the next window parameters. 
We have also presented a comprehensive analysis of empirical and statistical performance. When Table ~\ref{tab:result-overall} and Table ~\ref{tab:ttest_score} are analyzed, it is observed that the proposed method performance is outstanding in 3 out of 4 crypto-currency presented, and its average score for ten crypto-currency is smallest as compared to other state-of-the-art methods hence we can conclude that more similarity exists between the actual closing prices and predicted closing prices when compared for different Crypto-currencies. The proposed method can be applied to various other prediction applications where its challenging to model unseen volatile data, such as short-term load monitoring, sales and revenue prediction, and predicting hate intensity for social media content.}
\section{Acknowledgement}
The authors are indebted for the support provided by the Infosys Center for Artificial Intelligence at Indraprastha Institute of Information Technology- Delhi,(IIIT Delhi).
\section{Conclusion}
\cblue{
The current study forecasts the prices of crypto-currencies; this is halfway to the goal. The final objective is to take trading positions (BUY / SELL / HOLD). But that is a very challenging problem where strategies come into play. Unfortunately successful investors do not reveal their strategy. Pedagogically it is a matured area in traditional financial markets \citep{molinero2021influence} where game theory is mainly used in arriving at the decision boundaries; however how effective they are in practice is not known. Currently, a combination of game theory and social network analysis is used for arriving at such decision boundaries \citep{molinero2021influence}. In future, the authors would like to see if such cues from stock trading that can be used for maximising returns in crypto-currency trading.}

\bibliography{bibliography}

\begin{thebibliography}{68}
\expandafter\ifx\csname natexlab\endcsname\relax\def\natexlab#1{#1}\fi
\providecommand{\url}[1]{\texttt{#1}}
\providecommand{\href}[2]{#2}
\providecommand{\path}[1]{#1}
\providecommand{\DOIprefix}{doi:}
\providecommand{\ArXivprefix}{arXiv:}
\providecommand{\URLprefix}{URL: }
\providecommand{\Pubmedprefix}{pmid:}
\providecommand{\doi}[1]{\href{http://dx.doi.org/#1}{\path{#1}}}
\providecommand{\Pubmed}[1]{\href{pmid:#1}{\path{#1}}}
\providecommand{\bibinfo}[2]{#2}
\ifx\xfnm\relax \def\xfnm[#1]{\unskip,\space#1}\fi
\bibitem[{Abu~Bakar \& Rosbi(2017)}]{abu2017autoregressive}
\bibinfo{author}{Abu~Bakar, N.}, \& \bibinfo{author}{Rosbi, S.}
  (\bibinfo{year}{2017}).
\newblock \bibinfo{title}{Autoregressive integrated moving average (arima)
  model for forecasting cryptocurrency exchange rate in high volatility
  environment: A new insight of bitcoin transaction}.
\newblock {\it \bibinfo{journal}{International Journal of Advanced Engineering
  Research and Science}\/},  {\it \bibinfo{volume}{4}\/},
  \bibinfo{pages}{130--137}.
\bibitem[{Andersen et~al.(2009)Andersen, Davis, Krei{\ss} \&
  Mikosch}]{andersen2009handbook}
\bibinfo{author}{Andersen, T.~G.}, \bibinfo{author}{Davis, R.~A.},
  \bibinfo{author}{Krei{\ss}, J.-P.}, \& \bibinfo{author}{Mikosch, T.~V.}
  (\bibinfo{year}{2009}).
\newblock {\it \bibinfo{title}{Handbook of Financial Time Series}\/}.
\newblock \bibinfo{publisher}{Springer Science \& Business Media}.
\bibitem[{Andrieu et~al.(2010)Andrieu, Doucet \&
  Holenstein}]{andrieu2010particle}
\bibinfo{author}{Andrieu, C.}, \bibinfo{author}{Doucet, A.}, \&
  \bibinfo{author}{Holenstein, R.} (\bibinfo{year}{2010}).
\newblock \bibinfo{title}{Particle markov chain monte carlo methods}.
\newblock {\it \bibinfo{journal}{Journal of the Royal Statistical Society:
  Series B (Statistical Methodology)}\/},  {\it \bibinfo{volume}{72}\/},
  \bibinfo{pages}{269--342}.
\bibitem[{Baek \& Kim(2018)}]{baek2018modaugnet}
\bibinfo{author}{Baek, Y.}, \& \bibinfo{author}{Kim, H.~Y.}
  (\bibinfo{year}{2018}).
\newblock \bibinfo{title}{Modaugnet: A new forecasting framework for stock
  market index value with an overfitting prevention lstm module and a
  prediction lstm module}.
\newblock {\it \bibinfo{journal}{Expert Systems with Applications}\/},  {\it
  \bibinfo{volume}{113}\/}, \bibinfo{pages}{457--480}.
\bibitem[{Catania et~al.(2018)Catania, Grassi \&
  Ravazzolo}]{catania2018predicting}
\bibinfo{author}{Catania, L.}, \bibinfo{author}{Grassi, S.}, \&
  \bibinfo{author}{Ravazzolo, F.} (\bibinfo{year}{2018}).
\newblock \bibinfo{title}{Predicting the volatility of cryptocurrency
  time-series}.
\newblock {\it \bibinfo{journal}{Mathematical and Statistical Methods for
  Actuarial Sciences and Finance}\/},  (pp. \bibinfo{pages}{203--207}).
\bibitem[{Chen et~al.(2019)Chen, Jiang, Liao \& Zhao}]{chen2019efficient}
\bibinfo{author}{Chen, M.}, \bibinfo{author}{Jiang, H.}, \bibinfo{author}{Liao,
  W.}, \& \bibinfo{author}{Zhao, T.} (\bibinfo{year}{2019}).
\newblock \bibinfo{title}{Efficient approximation of deep relu networks for
  functions on low dimensional manifolds}.
\newblock {\it \bibinfo{journal}{Advances in Neural Information Processing
  Systems}\/},  {\it \bibinfo{volume}{32}\/}.
\bibitem[{Chen et~al.(2021)Chen, Jin, Liu \& Zhang}]{Chen2021deepNMF}
\bibinfo{author}{Chen, Z.}, \bibinfo{author}{Jin, S.}, \bibinfo{author}{Liu,
  R.}, \& \bibinfo{author}{Zhang, J.} (\bibinfo{year}{2021}).
\newblock \bibinfo{title}{A deep non-negative matrix factorization model for
  big data representation learning}.
\newblock {\it \bibinfo{journal}{Frontiers in Neurorobotics}\/},  {\it
  \bibinfo{volume}{15}\/}. \URLprefix
  \url{https://www.frontiersin.org/article/10.3389/fnbot.2021.701194}.
  \DOIprefix\doi{10.3389/fnbot.2021.701194}.
\bibitem[{Chopin et~al.(2013)Chopin, Jacob \&
  Papaspiliopoulos}]{chopin2013smc2}
\bibinfo{author}{Chopin, N.}, \bibinfo{author}{Jacob, P.~E.}, \&
  \bibinfo{author}{Papaspiliopoulos, O.} (\bibinfo{year}{2013}).
\newblock \bibinfo{title}{{SMC2}: an efficient algorithm for sequential
  analysis of state space models}.
\newblock {\it \bibinfo{journal}{Journal of the Royal Statistical Society:
  Series B (Statistical Methodology)}\/},  {\it \bibinfo{volume}{75}\/},
  \bibinfo{pages}{397--426}.
\bibitem[{Chouzenoux \& Elvira(2023)}]{chouzenoux2023graphit}
\bibinfo{author}{Chouzenoux, E.}, \& \bibinfo{author}{Elvira, V.}
  (\bibinfo{year}{2023}).
\newblock \bibinfo{title}{Graphit: Iterative reweighted l1 algorithm for sparse
  graph inference in state-space models}.
\newblock In {\it \bibinfo{booktitle}{ICASSP 2023-2023 IEEE International
  Conference on Acoustics, Speech and Signal Processing (ICASSP)}\/} (pp.
  \bibinfo{pages}{1--5}).
\newblock \bibinfo{organization}{IEEE}.
\bibitem[{Chouzenoux et~al.(2016)Chouzenoux, Pesquet \&
  Repetti}]{Chouzenoux2016}
\bibinfo{author}{Chouzenoux, E.}, \bibinfo{author}{Pesquet, J.-C.}, \&
  \bibinfo{author}{Repetti, A.} (\bibinfo{year}{2016}).
\newblock \bibinfo{title}{A block coordinate variable metric forward-backward
  algorithm}.
\newblock {\it \bibinfo{journal}{Journal of Global Optimization}\/},  {\it
  \bibinfo{volume}{66}\/}, \bibinfo{pages}{457--485}.
\bibitem[{Cox \& Elvira(2023)}]{cox2023sparse}
\bibinfo{author}{Cox, B.}, \& \bibinfo{author}{Elvira, V.}
  (\bibinfo{year}{2023}).
\newblock \bibinfo{title}{Sparse {B}ayesian estimation of parameters in
  linear-gaussian state-space models}.
\newblock {\it \bibinfo{journal}{IEEE Transactions on Signal Processing (to
  appear in)}\/}, .
\bibitem[{Crisan \& Miguez(2018)}]{crisan2018nested}
\bibinfo{author}{Crisan, D.}, \& \bibinfo{author}{Miguez, J.}
  (\bibinfo{year}{2018}).
\newblock \bibinfo{title}{Nested particle filters for online parameter
  estimation in discrete-time state-space markov models}.
\newblock {\it \bibinfo{journal}{Bernoulli}\/},  {\it \bibinfo{volume}{24}\/},
  \bibinfo{pages}{3039--3086}.
\bibitem[{Daubechies et~al.(2022)Daubechies, DeVore, Foucart, Hanin \&
  Petrova}]{daubechies2022nonlinear}
\bibinfo{author}{Daubechies, I.}, \bibinfo{author}{DeVore, R.},
  \bibinfo{author}{Foucart, S.}, \bibinfo{author}{Hanin, B.}, \&
  \bibinfo{author}{Petrova, G.} (\bibinfo{year}{2022}).
\newblock \bibinfo{title}{Nonlinear approximation and (deep) relu networks}.
\newblock {\it \bibinfo{journal}{Constructive Approximation}\/},  {\it
  \bibinfo{volume}{55}\/}, \bibinfo{pages}{127--172}.
\bibitem[{De~Handschutter et~al.(2021)De~Handschutter, Gillis \&
  Siebert}]{de2021survey}
\bibinfo{author}{De~Handschutter, P.}, \bibinfo{author}{Gillis, N.}, \&
  \bibinfo{author}{Siebert, X.} (\bibinfo{year}{2021}).
\newblock \bibinfo{title}{A survey on deep matrix factorizations}.
\newblock {\it \bibinfo{journal}{Computer Science Review}\/},  {\it
  \bibinfo{volume}{42}\/}, \bibinfo{pages}{100423}.
\bibitem[{Derbentsev et~al.(2020)Derbentsev, Matviychuk \&
  Soloviev}]{derbentsev2020forecasting}
\bibinfo{author}{Derbentsev, V.}, \bibinfo{author}{Matviychuk, A.}, \&
  \bibinfo{author}{Soloviev, V.~N.} (\bibinfo{year}{2020}).
\newblock \bibinfo{title}{Forecasting of cryptocurrency prices using machine
  learning}.
\newblock In {\it \bibinfo{booktitle}{Advanced Studies of Financial
  Technologies and Cryptocurrency Markets}\/} (pp. \bibinfo{pages}{211--231}).
\newblock \bibinfo{publisher}{Springer}.
\bibitem[{Digalakis et~al.(1993)Digalakis, Rohlicek \&
  Ostendorf}]{digalakis1993ml}
\bibinfo{author}{Digalakis, V.}, \bibinfo{author}{Rohlicek, J.~R.}, \&
  \bibinfo{author}{Ostendorf, M.} (\bibinfo{year}{1993}).
\newblock \bibinfo{title}{{ML} estimation of a stochastic linear system with
  the {EM} algorithm and its application to speech recognition}.
\newblock {\it \bibinfo{journal}{IEEE Transactions on Speech and Audio
  Processing}\/},  {\it \bibinfo{volume}{1}\/}, \bibinfo{pages}{431--442}.
\bibitem[{Dritsaki(2015)}]{dritsaki2015box}
\bibinfo{author}{Dritsaki, C.} (\bibinfo{year}{2015}).
\newblock \bibinfo{title}{Box-{J}enkins modeling of {G}reek stock prices data}.
\newblock {\it \bibinfo{journal}{International Journal of Economics and
  Financial Issues}\/},  {\it \bibinfo{volume}{5}\/}.
\bibitem[{Elbr{\"a}chter et~al.(2021)Elbr{\"a}chter, Perekrestenko, Grohs \&
  B{\"o}lcskei}]{elbrachter2021deep}
\bibinfo{author}{Elbr{\"a}chter, D.}, \bibinfo{author}{Perekrestenko, D.},
  \bibinfo{author}{Grohs, P.}, \& \bibinfo{author}{B{\"o}lcskei, H.}
  (\bibinfo{year}{2021}).
\newblock \bibinfo{title}{Deep neural network approximation theory}.
\newblock {\it \bibinfo{journal}{IEEE Transactions on Information Theory}\/},
  {\it \bibinfo{volume}{67}\/}, \bibinfo{pages}{2581--2623}.
\bibitem[{Elsworth \& G{\"u}ttel(2020)}]{elsworth2020time}
\bibinfo{author}{Elsworth, S.}, \& \bibinfo{author}{G{\"u}ttel, S.}
  (\bibinfo{year}{2020}).
\newblock \bibinfo{title}{Time series forecasting using {LSTM} networks: A
  symbolic approach}.
\newblock {\it \bibinfo{journal}{\url{https://arxiv.org/abs/2003.05672}}\/}, .
\bibitem[{Elvira \& Chouzenoux(2022)}]{ElviraGraphEM2022}
\bibinfo{author}{Elvira, V.}, \& \bibinfo{author}{Chouzenoux, E.}
  (\bibinfo{year}{2022}).
\newblock \bibinfo{title}{Graphical inference in linear-{G}aussian state-space
  models}.
\newblock {\it \bibinfo{journal}{IEEE Transactions on Signal Processing}\/},
  {\it \bibinfo{volume}{70}\/}, \bibinfo{pages}{4757--4771}.
\bibitem[{Elvira et~al.(2017)Elvira, M{\'\i}guez \&
  Djuri{\'c}}]{elvira2017adapting}
\bibinfo{author}{Elvira, V.}, \bibinfo{author}{M{\'\i}guez, J.}, \&
  \bibinfo{author}{Djuri{\'c}, P.~M.} (\bibinfo{year}{2017}).
\newblock \bibinfo{title}{Adapting the number of particles in sequential monte
  carlo methods through an online scheme for convergence assessment}.
\newblock {\it \bibinfo{journal}{IEEE Transactions on Signal Processing}\/},
  {\it \bibinfo{volume}{65}\/}, \bibinfo{pages}{1781--1794}.
\bibitem[{Flenner \& Hunter(2017)}]{flenner2017deep}
\bibinfo{author}{Flenner, J.}, \& \bibinfo{author}{Hunter, B.}
  (\bibinfo{year}{2017}).
\newblock \bibinfo{title}{A deep non-negative matrix factorization neural
  network}.
\newblock {\it \bibinfo{journal}{Semantic Scholar}\/}, .
\bibitem[{Garzon \& Botelho(1999)}]{garzon1999dynamical}
\bibinfo{author}{Garzon, M.}, \& \bibinfo{author}{Botelho, F.}
  (\bibinfo{year}{1999}).
\newblock \bibinfo{title}{Dynamical approximation by recurrent neural
  networks}.
\newblock {\it \bibinfo{journal}{Neurocomputing}\/},  {\it
  \bibinfo{volume}{29}\/}, \bibinfo{pages}{25--46}.
\bibitem[{Glenski et~al.(2019)Glenski, Weninger \&
  Volkova}]{glenski2019improved}
\bibinfo{author}{Glenski, M.}, \bibinfo{author}{Weninger, T.}, \&
  \bibinfo{author}{Volkova, S.} (\bibinfo{year}{2019}).
\newblock \bibinfo{title}{Improved forecasting of cryptocurrency price using
  social signals}.
\newblock {\it \bibinfo{journal}{\url{https://arxiv.org/abs/1907.00558}}\/}, .
\bibitem[{Hammer(2000)}]{BarbaraHammer}
\bibinfo{author}{Hammer, B.} (\bibinfo{year}{2000}).
\newblock \bibinfo{title}{On the approximation capability of recurrent neural
  networks}.
\newblock {\it \bibinfo{journal}{Neurocomputing}\/},  {\it
  \bibinfo{volume}{31}\/}, \bibinfo{pages}{107--123}.
\bibitem[{Jacobson \& Fessler(2007)}]{jacobson}
\bibinfo{author}{Jacobson, M.}, \& \bibinfo{author}{Fessler, J.}
  (\bibinfo{year}{2007}).
\newblock \bibinfo{title}{An expanded theoretical treatment of
  iteration-dependent majorize-minimize algorithms}.
\newblock {\it \bibinfo{journal}{IEEE Transactions on Image Processing}\/},
  {\it \bibinfo{volume}{16}\/}, \bibinfo{pages}{2411–2422}.
\bibitem[{Jarrett \& Kyper(2011)}]{jarrett2011arima}
\bibinfo{author}{Jarrett, J.~E.}, \& \bibinfo{author}{Kyper, E.}
  (\bibinfo{year}{2011}).
\newblock \bibinfo{title}{Arima modeling with intervention to forecast and
  analyze chinese stock prices}.
\newblock {\it \bibinfo{journal}{International Journal of Engineering Business
  Management}\/},  {\it \bibinfo{volume}{3}\/}, \bibinfo{pages}{53--58}.
\bibitem[{Kim et~al.(2021)Kim, Trimborn \& H{\"a}rdle}]{kim2021vcrix}
\bibinfo{author}{Kim, A.}, \bibinfo{author}{Trimborn, S.}, \&
  \bibinfo{author}{H{\"a}rdle, W.~K.} (\bibinfo{year}{2021}).
\newblock \bibinfo{title}{Vcrix—a volatility index for crypto-currencies}.
\newblock {\it \bibinfo{journal}{International Review of Financial
  Analysis}\/},  {\it \bibinfo{volume}{78}\/}, \bibinfo{pages}{101915}.
\bibitem[{Kim \& Kim(2019)}]{kim2019forecasting}
\bibinfo{author}{Kim, T.}, \& \bibinfo{author}{Kim, H.~Y.}
  (\bibinfo{year}{2019}).
\newblock \bibinfo{title}{Forecasting stock prices with a feature fusion
  lstm-cnn model using different representations of the same data}.
\newblock {\it \bibinfo{journal}{PloS one}\/},  {\it \bibinfo{volume}{14}\/},
  \bibinfo{pages}{e0212320}.
\bibitem[{K{\"o}chling et~al.(2020)K{\"o}chling, Schmidtke \&
  Posch}]{kochling2020volatility}
\bibinfo{author}{K{\"o}chling, G.}, \bibinfo{author}{Schmidtke, P.}, \&
  \bibinfo{author}{Posch, P.~N.} (\bibinfo{year}{2020}).
\newblock \bibinfo{title}{Volatility forecasting accuracy for bitcoin}.
\newblock {\it \bibinfo{journal}{Economics Letters}\/},  {\it
  \bibinfo{volume}{191}\/}, \bibinfo{pages}{108836}.
\bibitem[{Kraaijeveld \& De~Smedt(2020)}]{kraaijeveld2020predictive}
\bibinfo{author}{Kraaijeveld, O.}, \& \bibinfo{author}{De~Smedt, J.}
  (\bibinfo{year}{2020}).
\newblock \bibinfo{title}{The predictive power of public twitter sentiment for
  forecasting cryptocurrency prices}.
\newblock {\it \bibinfo{journal}{Journal of International Financial Markets,
  Institutions and Money}\/},  {\it \bibinfo{volume}{65}\/},
  \bibinfo{pages}{101188}.
\bibitem[{Kristjanpoller \& Minutolo(2018)}]{kristjanpoller2018hybrid}
\bibinfo{author}{Kristjanpoller, W.}, \& \bibinfo{author}{Minutolo, M.~C.}
  (\bibinfo{year}{2018}).
\newblock \bibinfo{title}{A hybrid volatility forecasting framework integrating
  garch, artificial neural network, technical analysis and principal components
  analysis}.
\newblock {\it \bibinfo{journal}{Expert Systems with Applications}\/},  {\it
  \bibinfo{volume}{109}\/}, \bibinfo{pages}{1--11}.
\bibitem[{Liang \& Srikant(2016)}]{liang2016deep}
\bibinfo{author}{Liang, S.}, \& \bibinfo{author}{Srikant, R.}
  (\bibinfo{year}{2016}).
\newblock \bibinfo{title}{Why deep neural networks for function approximation?}
\newblock {\it \bibinfo{journal}{arXiv preprint arXiv:1610.04161}\/}, .
\bibitem[{Lim et~al.(2019)Lim, Arik, Loeff \& Pfister}]{tft}
\bibinfo{author}{Lim, B.}, \bibinfo{author}{Arik, S.~{\"{O}}.},
  \bibinfo{author}{Loeff, N.}, \& \bibinfo{author}{Pfister, T.}
  (\bibinfo{year}{2019}).
\newblock \bibinfo{title}{Temporal fusion transformers for interpretable
  multi-horizon time series forecasting}.
\newblock {\it \bibinfo{journal}{CoRR}\/},  {\it
  \bibinfo{volume}{abs/1912.09363}\/}. \URLprefix
  \url{http://arxiv.org/abs/1912.09363}.
  \href{http://arxiv.org/abs/1912.09363}{\tt arXiv:1912.09363}.
\bibitem[{Lin et~al.(2015)Lin, Ma \& Zhang}]{lin2015global}
\bibinfo{author}{Lin, T.}, \bibinfo{author}{Ma, S.}, \& \bibinfo{author}{Zhang,
  S.} (\bibinfo{year}{2015}).
\newblock \bibinfo{title}{On the global linear convergence of the admm with
  multiblock variables}.
\newblock {\it \bibinfo{journal}{SIAM Journal on Optimization}\/},  {\it
  \bibinfo{volume}{25}\/}, \bibinfo{pages}{1478--1497}.
\bibitem[{Liu \& Liang(2021)}]{liu2021optimal}
\bibinfo{author}{Liu, B.}, \& \bibinfo{author}{Liang, Y.}
  (\bibinfo{year}{2021}).
\newblock \bibinfo{title}{Optimal function approximation with relu neural
  networks}.
\newblock {\it \bibinfo{journal}{Neurocomputing}\/},  {\it
  \bibinfo{volume}{435}\/}, \bibinfo{pages}{216--227}.
\bibitem[{Livieris et~al.(2021)Livieris, Kiriakidou, Stavroyiannis \&
  Pintelas}]{livieris2021advanced}
\bibinfo{author}{Livieris, I.~E.}, \bibinfo{author}{Kiriakidou, N.},
  \bibinfo{author}{Stavroyiannis, S.}, \& \bibinfo{author}{Pintelas, P.}
  (\bibinfo{year}{2021}).
\newblock \bibinfo{title}{An advanced cnn-lstm model for cryptocurrency
  forecasting}.
\newblock {\it \bibinfo{journal}{Electronics}\/},  {\it
  \bibinfo{volume}{10}\/}, \bibinfo{pages}{287}.
\bibitem[{Long et~al.(2019)Long, Lu \& Cui}]{long2019deep}
\bibinfo{author}{Long, W.}, \bibinfo{author}{Lu, Z.}, \& \bibinfo{author}{Cui,
  L.} (\bibinfo{year}{2019}).
\newblock \bibinfo{title}{Deep learning-based feature engineering for stock
  price movement prediction}.
\newblock {\it \bibinfo{journal}{Knowledge-Based Systems}\/},  {\it
  \bibinfo{volume}{164}\/}, \bibinfo{pages}{163--173}.
\bibitem[{Ma et~al.(2020{\natexlab{a}})Ma, Liang, Ma \&
  Wahab}]{ma2020cryptocurrency}
\bibinfo{author}{Ma, F.}, \bibinfo{author}{Liang, C.}, \bibinfo{author}{Ma,
  Y.}, \& \bibinfo{author}{Wahab, M.} (\bibinfo{year}{2020}{\natexlab{a}}).
\newblock \bibinfo{title}{Cryptocurrency volatility forecasting: A markov
  regime-switching midas approach}.
\newblock {\it \bibinfo{journal}{Journal of Forecasting}\/},  {\it
  \bibinfo{volume}{39}\/}, \bibinfo{pages}{1277--1290}.
\bibitem[{Ma et~al.(2020{\natexlab{b}})Ma, Karkus, Hsu \& Lee}]{ma2020particle}
\bibinfo{author}{Ma, X.}, \bibinfo{author}{Karkus, P.}, \bibinfo{author}{Hsu,
  D.}, \& \bibinfo{author}{Lee, W.~S.} (\bibinfo{year}{2020}{\natexlab{b}}).
\newblock \bibinfo{title}{Particle filter recurrent neural networks}.
\newblock In {\it \bibinfo{booktitle}{Proceedings of the AAAI Conference on
  Artificial Intelligence}\/} (pp. \bibinfo{pages}{5101--5108}).
\newblock volume~\bibinfo{volume}{34}.
\bibitem[{Mahdizadehaghdam et~al.(2019)Mahdizadehaghdam, Panahi, Krim \&
  Dai}]{mahdizadehaghdam2019deep}
\bibinfo{author}{Mahdizadehaghdam, S.}, \bibinfo{author}{Panahi, A.},
  \bibinfo{author}{Krim, H.}, \& \bibinfo{author}{Dai, L.}
  (\bibinfo{year}{2019}).
\newblock \bibinfo{title}{Deep dictionary learning: A parametric network
  approach}.
\newblock {\it \bibinfo{journal}{IEEE Transactions on Image Processing}\/},
  {\it \bibinfo{volume}{28}\/}, \bibinfo{pages}{4790--4802}.
\bibitem[{Mei et~al.(2019)Mei, De~Castro, Goude, Azaïs \& Hébrail}]{Mei}
\bibinfo{author}{Mei, J.}, \bibinfo{author}{De~Castro, Y.},
  \bibinfo{author}{Goude, Y.}, \bibinfo{author}{Azaïs, J.-M.}, \&
  \bibinfo{author}{Hébrail, G.} (\bibinfo{year}{2019}).
\newblock \bibinfo{title}{Nonnegative matrix factorization with side
  information for time series recovery and prediction}.
\newblock {\it \bibinfo{journal}{IEEE Transactions on Knowledge and Data
  Engineering}\/},  {\it \bibinfo{volume}{31}\/}, \bibinfo{pages}{493--506}.
  \DOIprefix\doi{10.1109/TKDE.2018.2839678}.
\bibitem[{Molinero \& Riquelme(2021)}]{molinero2021influence}
\bibinfo{author}{Molinero, X.}, \& \bibinfo{author}{Riquelme, F.}
  (\bibinfo{year}{2021}).
\newblock \bibinfo{title}{Influence decision models: from cooperative game
  theory to social network analysis}.
\newblock {\it \bibinfo{journal}{Computer Science Review}\/},  {\it
  \bibinfo{volume}{39}\/}, \bibinfo{pages}{100343}.
\bibitem[{Montella(2011)}]{montella2011kalman}
\bibinfo{author}{Montella, C.} (\bibinfo{year}{2011}).
\newblock \bibinfo{title}{The kalman filter and related algorithms: A
  literature review}.
\newblock {\it \bibinfo{journal}{Res. Gate}\/},  (pp. \bibinfo{pages}{1--17}).
\bibitem[{Nayak et~al.(2021)Nayak, Nayak \& Das}]{nayak2021modeling}
\bibinfo{author}{Nayak, S.~K.}, \bibinfo{author}{Nayak, S.~C.}, \&
  \bibinfo{author}{Das, S.} (\bibinfo{year}{2021}).
\newblock \bibinfo{title}{Modeling and forecasting cryptocurrency closing
  prices with rao algorithm-based artificial neural networks: A machine
  learning approach}.
\newblock {\it \bibinfo{journal}{FinTech}\/},  {\it \bibinfo{volume}{1}\/},
  \bibinfo{pages}{47--62}.
\bibitem[{Newman et~al.(2023)Newman, King, Elvira, de~Valpine, McCrea \&
  Morgan}]{newman2023state}
\bibinfo{author}{Newman, K.}, \bibinfo{author}{King, R.},
  \bibinfo{author}{Elvira, V.}, \bibinfo{author}{de~Valpine, P.},
  \bibinfo{author}{McCrea, R.~S.}, \& \bibinfo{author}{Morgan, B.~J.}
  (\bibinfo{year}{2023}).
\newblock \bibinfo{title}{State-space models for ecological time-series data:
  Practical model-fitting}.
\newblock {\it \bibinfo{journal}{Methods in Ecology and Evolution}\/},  {\it
  \bibinfo{volume}{14}\/}, \bibinfo{pages}{26--42}.
\bibitem[{Nishihara et~al.(2015)Nishihara, Lessard, Recht, Packard \&
  Jordan}]{nishihara2015general}
\bibinfo{author}{Nishihara, R.}, \bibinfo{author}{Lessard, L.},
  \bibinfo{author}{Recht, B.}, \bibinfo{author}{Packard, A.}, \&
  \bibinfo{author}{Jordan, M.} (\bibinfo{year}{2015}).
\newblock \bibinfo{title}{A general analysis of the convergence of admm}.
\newblock In {\it \bibinfo{booktitle}{International Conference on Machine
  Learning}\/} (pp. \bibinfo{pages}{343--352}).
\newblock \bibinfo{organization}{PMLR}.
\bibitem[{Oreshkin et~al.(2019)Oreshkin, Carpov, Chapados \&
  Bengio}]{oreshkin2019n}
\bibinfo{author}{Oreshkin, B.~N.}, \bibinfo{author}{Carpov, D.},
  \bibinfo{author}{Chapados, N.}, \& \bibinfo{author}{Bengio, Y.}
  (\bibinfo{year}{2019}).
\newblock \bibinfo{title}{N-beats: Neural basis expansion analysis for
  interpretable time series forecasting}.
\newblock {\it \bibinfo{journal}{arXiv preprint arXiv:1905.10437}\/}, .
\bibitem[{{R. Molla}(2021)}]{molla2021elon}
\bibinfo{author}{{R. Molla}} (\bibinfo{year}{2021}).
\newblock \bibinfo{title}{When {Elon Musk} tweets, crypto prices move}.
\newblock {\it
  \bibinfo{journal}{\url{https://www.vox.com/recode/2021/5/18/22441831/elon-musk-bitcoin-dogecoin-crypto-prices-tesla}}\/},
  .
\bibitem[{Rangapuram et~al.(2018)Rangapuram, Seeger, Gasthaus, Stella, Wang \&
  Januschowski}]{rangapuram2018deep}
\bibinfo{author}{Rangapuram, S.~S.}, \bibinfo{author}{Seeger, M.~W.},
  \bibinfo{author}{Gasthaus, J.}, \bibinfo{author}{Stella, L.},
  \bibinfo{author}{Wang, Y.}, \& \bibinfo{author}{Januschowski, T.}
  (\bibinfo{year}{2018}).
\newblock \bibinfo{title}{Deep state space models for time series forecasting}.
\newblock {\it \bibinfo{journal}{Advances in Neural Information Processing
  Systems}\/},  {\it \bibinfo{volume}{31}\/}, \bibinfo{pages}{7785--7794}.
\bibitem[{Rankin(1986)}]{kalmanthesis}
\bibinfo{author}{Rankin, J.} (\bibinfo{year}{1986}).
\newblock {\it \bibinfo{title}{{ Kalman filtering approach to market price
  forecasting }}\/}.
\newblock \bibinfo{type}{{Ph.D.} diss.} Iowa State University.
\bibitem[{Rounaghi \& Zadeh(2016)}]{rounaghi2016investigation}
\bibinfo{author}{Rounaghi, M.~M.}, \& \bibinfo{author}{Zadeh, F.~N.}
  (\bibinfo{year}{2016}).
\newblock \bibinfo{title}{Investigation of market efficiency and financial
  stability between s\&p 500 and london stock exchange: monthly and yearly
  forecasting of time series stock returns using arma model}.
\newblock {\it \bibinfo{journal}{Physica A: Statistical Mechanics and its
  Applications}\/},  {\it \bibinfo{volume}{456}\/}, \bibinfo{pages}{10--21}.
\bibitem[{{S. Soni}(2021)}]{Financial_express}
\bibinfo{author}{{S. Soni}} (\bibinfo{year}{2021}).
\newblock \bibinfo{title}{Crypto investors lost \$748 billion in last seven
  days as bitcoin, ethereum, dogecoin, others declined}.
\newblock
  \bibinfo{howpublished}{\url{https://www.financialexpress.com/market/crypto-investors-lost-748-billion-in-last/seven-days-as-bitcoin-ethereum-dogecoin-others-declined}}.
\newblock \bibinfo{note}{Accessed: 2021-05-23}.
\bibitem[{Salinas et~al.(2020)Salinas, Flunkert, Gasthaus \&
  Januschowski}]{salinas2020deepar}
\bibinfo{author}{Salinas, D.}, \bibinfo{author}{Flunkert, V.},
  \bibinfo{author}{Gasthaus, J.}, \& \bibinfo{author}{Januschowski, T.}
  (\bibinfo{year}{2020}).
\newblock \bibinfo{title}{Deepar: Probabilistic forecasting with autoregressive
  recurrent networks}.
\newblock {\it \bibinfo{journal}{International Journal of Forecasting}\/},
  {\it \bibinfo{volume}{36}\/}, \bibinfo{pages}{1181--1191}.
\bibitem[{S{\"a}rkk{\"a}(2013)}]{sarkka2013bayesian}
\bibinfo{author}{S{\"a}rkk{\"a}, S.} (\bibinfo{year}{2013}).
\newblock {\it \bibinfo{title}{Bayesian Filtering and Smoothing}\/}.
\newblock (\bibinfo{edition}{3rd} ed.).
\newblock \bibinfo{publisher}{Cambridge University Press}.
\bibitem[{Sezer \& Ozbayoglu(2018)}]{sezer2018algorithmic}
\bibinfo{author}{Sezer, O.~B.}, \& \bibinfo{author}{Ozbayoglu, A.~M.}
  (\bibinfo{year}{2018}).
\newblock \bibinfo{title}{Algorithmic financial trading with deep convolutional
  neural networks: Time series to image conversion approach}.
\newblock {\it \bibinfo{journal}{Applied Soft Computing}\/},  {\it
  \bibinfo{volume}{70}\/}, \bibinfo{pages}{525--538}.
\bibitem[{Sharma et~al.(2021)Sharma, Elvira, Chouzenoux \&
  Majumdar}]{sharma2021recurrent}
\bibinfo{author}{Sharma, S.}, \bibinfo{author}{Elvira, V.},
  \bibinfo{author}{Chouzenoux, E.}, \& \bibinfo{author}{Majumdar, A.}
  (\bibinfo{year}{2021}).
\newblock \bibinfo{title}{Recurrent dictionary learning for state-space models
  with an application in stock forecasting}.
\newblock {\it \bibinfo{journal}{Neurocomputing}\/},  {\it
  \bibinfo{volume}{450}\/}, \bibinfo{pages}{1--13}.
\bibitem[{Sharma \& Majumdar(2021)}]{sharma2021sequential}
\bibinfo{author}{Sharma, S.}, \& \bibinfo{author}{Majumdar, A.}
  (\bibinfo{year}{2021}).
\newblock \bibinfo{title}{Sequential transform learning}.
\newblock {\it \bibinfo{journal}{ACM Transactions on Knowledge Discovery from
  Data (TKDD)}\/},  {\it \bibinfo{volume}{15}\/}, \bibinfo{pages}{1--18}.
\bibitem[{Sharma et~al.(2020)Sharma, Majumdar, Elvira \&
  Chouzenoux}]{sharma2020blind}
\bibinfo{author}{Sharma, S.}, \bibinfo{author}{Majumdar, A.},
  \bibinfo{author}{Elvira, V.}, \& \bibinfo{author}{Chouzenoux, E.}
  (\bibinfo{year}{2020}).
\newblock \bibinfo{title}{Blind kalman filtering for short-term load
  forecasting}.
\newblock {\it \bibinfo{journal}{IEEE Transactions on Power Systems}\/},  {\it
  \bibinfo{volume}{35}\/}, \bibinfo{pages}{4916--4919}.
\bibitem[{Shumway \& Stoffer(1982)}]{shumway1982approach}
\bibinfo{author}{Shumway, R.~H.}, \& \bibinfo{author}{Stoffer, D.~S.}
  (\bibinfo{year}{1982}).
\newblock \bibinfo{title}{An approach to time series smoothing and forecasting
  using the {EM} algorithm}.
\newblock {\it \bibinfo{journal}{Journal of Time Series Analysis}\/},  {\it
  \bibinfo{volume}{3}\/}, \bibinfo{pages}{253--264}.
\bibitem[{Tariyal et~al.(2016)Tariyal, Majumdar, Singh \&
  Vatsa}]{tariyal2016deep}
\bibinfo{author}{Tariyal, S.}, \bibinfo{author}{Majumdar, A.},
  \bibinfo{author}{Singh, R.}, \& \bibinfo{author}{Vatsa, M.}
  (\bibinfo{year}{2016}).
\newblock \bibinfo{title}{Deep dictionary learning}.
\newblock {\it \bibinfo{journal}{IEEE Access}\/},  {\it \bibinfo{volume}{4}\/},
  \bibinfo{pages}{10096--10109}.
\bibitem[{Trigeorgis et~al.(2016)Trigeorgis, Bousmalis, Zafeiriou \&
  Schuller}]{trigeorgis2016deep}
\bibinfo{author}{Trigeorgis, G.}, \bibinfo{author}{Bousmalis, K.},
  \bibinfo{author}{Zafeiriou, S.}, \& \bibinfo{author}{Schuller, B.~W.}
  (\bibinfo{year}{2016}).
\newblock \bibinfo{title}{A deep matrix factorization method for learning
  attribute representations}.
\newblock {\it \bibinfo{journal}{IEEE Transactions on Pattern Analysis and
  Machine Intelligence}\/},  {\it \bibinfo{volume}{39}\/},
  \bibinfo{pages}{417--429}.
\bibitem[{Walther et~al.(2019)Walther, Klein \& Bouri}]{walther2019exogenous}
\bibinfo{author}{Walther, T.}, \bibinfo{author}{Klein, T.}, \&
  \bibinfo{author}{Bouri, E.} (\bibinfo{year}{2019}).
\newblock \bibinfo{title}{Exogenous drivers of bitcoin and cryptocurrency
  volatility--a mixed data sampling approach to forecasting}.
\newblock {\it \bibinfo{journal}{Journal of International Financial Markets,
  Institutions and Money}\/},  {\it \bibinfo{volume}{63}\/},
  \bibinfo{pages}{101133}.
\bibitem[{Wang et~al.(2019)Wang, Yin \& Zeng}]{wang2019global}
\bibinfo{author}{Wang, Y.}, \bibinfo{author}{Yin, W.}, \&
  \bibinfo{author}{Zeng, J.} (\bibinfo{year}{2019}).
\newblock \bibinfo{title}{Global convergence of admm in nonconvex nonsmooth
  optimization}.
\newblock {\it \bibinfo{journal}{Journal of Scientific Computing}\/},  {\it
  \bibinfo{volume}{78}\/}, \bibinfo{pages}{29--63}.
\bibitem[{Woebbeking(2021)}]{woebbeking2021cryptocurrency}
\bibinfo{author}{Woebbeking, F.} (\bibinfo{year}{2021}).
\newblock \bibinfo{title}{Cryptocurrency volatility markets}.
\newblock {\it \bibinfo{journal}{Digital Finance}\/},  {\it
  \bibinfo{volume}{3}\/}, \bibinfo{pages}{273--298}.
\bibitem[{Yarotsky(2018)}]{yarotsky2018optimal}
\bibinfo{author}{Yarotsky, D.} (\bibinfo{year}{2018}).
\newblock \bibinfo{title}{Optimal approximation of continuous functions by very
  deep relu networks}.
\newblock In {\it \bibinfo{booktitle}{Proceedings of the Conference on Learning
  Theory (COLT 2018)}\/} (pp. \bibinfo{pages}{639--649}).
\newblock \bibinfo{organization}{PMLR}.
\bibitem[{Yasir et~al.(2020)Yasir, Attique, Latif, Chaudhary, Afzal, Ahmed \&
  Shahzad}]{yasir2020deep}
\bibinfo{author}{Yasir, M.}, \bibinfo{author}{Attique, M.},
  \bibinfo{author}{Latif, K.}, \bibinfo{author}{Chaudhary, G.~M.},
  \bibinfo{author}{Afzal, S.}, \bibinfo{author}{Ahmed, K.}, \&
  \bibinfo{author}{Shahzad, F.} (\bibinfo{year}{2020}).
\newblock \bibinfo{title}{Deep-learning-assisted business intelligence model
  for cryptocurrency forecasting using social media sentiment}.
\newblock {\it \bibinfo{journal}{Journal of Enterprise Information
  Management}\/}, .
\bibitem[{Ye \& Dai(2022)}]{ye2022relationship}
\bibinfo{author}{Ye, R.}, \& \bibinfo{author}{Dai, Q.} (\bibinfo{year}{2022}).
\newblock \bibinfo{title}{A relationship-aligned transfer learning algorithm
  for time series forecasting}.
\newblock {\it \bibinfo{journal}{Information Sciences}\/},  {\it
  \bibinfo{volume}{593}\/}, \bibinfo{pages}{17--34}.

\end{thebibliography}

\end{document}